\documentclass[aps,twocolumn,superscriptaddress,showpacs,showkeys]{revtex4}
\usepackage{graphics,graphicx,dcolumn,bm,fleqn,epic,eepic,float}
\usepackage{amssymb,amsmath,multirow,rotate,color,float,times}
\usepackage{color}
\usepackage{soul}                    
\definecolor{red}{rgb}{1,0,0}
\definecolor{green}{rgb}{0,1,0}
\definecolor{blue}{rgb}{0,0,1}

\newcommand{\p}{\partial}
\newcommand{\mbf}{\mathbf}
\newcommand{\bdif}{\,\boldsymbol{d}}
\newcommand{\fracpp}[1]{\frac{\p}{\p#1}}
\newcommand{\Ito}{It{\^o}}

\newcommand{\vecDdrift}{\mbf{D}^{(1)}}
\newcommand{\vecDdiff}{\mbf{D}^{(2)}}
\newcommand{\eps}{\varepsilon}

\newcommand{\vecxi}{\boldsymbol{\xi}}

\newcommand{\vecGamma}{\boldsymbol{\Gamma}}

\newcommand{\vecalpha}{{\boldsymbol{\alpha}}}
\newcommand{\vecbeta}{{\boldsymbol{\beta}}}

\newcommand{\vecc}{\mbf{c}}
\newcommand{\vecf}{\mbf{f}}

\newcommand{\vecR}{\mbf{R}}

\newcommand{\vecX}{\mbf{X}}
\newcommand{\vecy}{\mbf{y}}

\newcommand{\vecY}{\mbf{Y}}

\newcommand{\vecz}{\mbf{z}}

\newcommand{\matg}{\mbf{g}}

\newcommand{\matM}{\mbf{M}}

\newcommand{\matV}{\mbf{V}}

\newcommand{\mom}[1]{m^{(#1)}}
\newcommand{\bmom}[1]{\mbf{m}^{(#1)}}

\newcommand{\smom}[1]{m^{*(#1)}}
\newcommand{\bsmom}[1]{\mbf{m}^{*(#1)}}
\newcommand{\hmom}[1]{\hat m^{(#1)}}
\newcommand{\bhmom}[1]{\hat{\mbf{m}}^{(#1)}}

\begin{document}

\title{How to analyze stochastic time series obeying a 2nd order differential equation}

\author{B.~Lehle}
\author{J.~Peinke}
\affiliation{Institute of Physics, University of Oldenburg, D-2611 Oldenburg, Germany}

\begin{abstract}
The stochastic properties of a Langevin-type Markov process can be extracted from a given time series by a
Markov analysis. Also processes that obey a stochastically forced second order differential equation can be analyzed this way
by employing a particular embedding approach: To obtain a Markovian process in 2N dimensions from a non Markovian signal
in N dimensions, the system is described in a phase space that is extended by the temporal derivative of the signal.
For a discrete time series, however, this derivative can only be calculated by a differencing scheme, which introduces an error.
If the effects of this error are not accounted for, this leads to systematic errors in the estimation of the drift- and
diffusion functions of the process. In this paper we will analyze these errors and we will propose an approach
that correctly accounts for them. This approach allows an accurate parameter estimation and, additionally, is able to cope
with weak measurement noise, which may be superimposed to a given time series.
\end{abstract}




\maketitle

\section{Introduction}
\label{sec_intro}

\noindent Many dynamical systems can be modelled as continuous-time Markov processes $\vecY(t)$ that are driven by
Gaussian white noise $\vecxi(t)$ with $\left<\xi_i(t)\right>\!=\!0$ and
$\left<\xi_i(t)\xi_j(t')\right>\!=\!\delta_{ij}\delta(t\!-\!t')$. The temporal evolution of such a process obeys a
Langevin equation -- a first order ordinary differential equation (ODE) that is stochastically forced
\begin{eqnarray}\label{Langevin_Y}
\dot \vecY &=& \mbf{a}(\vecY)+\mbf{b}(\vecY)\,\vecxi(t).
\end{eqnarray}

\noindent Here and in the following \Ito's definition of a stochastic integral is used \cite{platen99}. Furthermore,
a stationary stochastic process is looked at, whereas in general $\mbf{a}$ and $\mbf{b}$ may depend on time.

The Kramers--Moyal coefficients of the Fokker--Planck equation corresponding to Eq.~(\ref{Langevin_Y}) are
denoted by $\vecDdrift$ and $\vecDdiff$ and commonly referred to as drift- and diffusion function respectively
\cite{risken89}. These functions uniquely define the stochastic process and are related to $\mbf{a}$ and $\mbf{b}$ by
\begin{eqnarray}
\vecDdrift(\vecy) &=& \mbf{a}(\vecy),\qquad
\vecDdiff(\vecy) \;=\; \mbf{b}(\vecy)\mbf{b}^t(\vecy).
\end{eqnarray}

\noindent It is possible to estimate $\vecDdrift$ and $\vecDdiff$ from a given time series of $\vecY$ by a Markov
analysis. This technique, also denoted as direct estimation method, has been introduced in the late 1990s
\cite{friedrich97,siegert98,friedrich00,gradisek00}. Since then it has been successfully applied to problems out of
many different fields. Reviews on Markov analysis and its applications can be found e.g. in \cite{friedrich11,friedrich08}.

The method is based on the fact that the moments $\mbf{M}^{(k)}$ of the conditional process increments
of $\vecY$ can be expressed in terms of the Kramers--Moyal coefficients
\begin{eqnarray}\label{moments_incY}
\mbf{M}^{(k)}(\vecy,\tau) &:=& \left<\big[\vecY(t\!+\!\tau)\!-\!\vecY(t)\big]^k\right>\big|_{\vecY(t)=\vecy}\cr
 &=& \tau\mbf{D}^{(k)}(\vecy)+O(\tau^2) ,\quad k\;=\; 1,2.
\end{eqnarray}

\noindent Here and in the following the $k$-th power of a vector denotes a $k$-fold dyadic product.
The time argument $t$ of $\mbf{M}^{(k)}$ is suppressed here because a stationary process is assumed.
This assumption also allows a moment estimation from a single time series -- ensemble averages can be replaced by
time averages then (tacitly assuming ergodicity). For a non-stationary process an ensemble of time series would be
needed (alternatively a windowing strategy could be applied, assuming a slowly varying time dependence).

The moments $\mbf{M}^{(k)}$ (Eq.~(\ref{moments_incY})) can be expressed in terms of moments
$\bmom{k}$ of the two-point probability density function (PDF) of $\vecY$ at times $t$ and $t\!+\!\tau$.
These moments $\bmom{k}$ are defined as
\begin{eqnarray}\label{def_momk}
\bmom{k}(\vecy,\tau) &:=& \int_{\mbf{s}}(\mbf{s}\!-\!\vecy)^k p(\vecy,t;\mbf{s},t\!+\!\tau)\bdif s,
\end{eqnarray}

\noindent where again the time argument $t$ is suppressed because of the assumption of stationarity.
Using the well known relations $p(a;b)\!=\!p(b)p(a|b)$ and $\int_a f(a)p(a|b)\!=\!\left<f(A)|b\right>$ leads to
\begin{eqnarray}\label{relation_m_h}
\bmom{k}(\vecy,\tau) &=& p(\vecy,t)\,\mbf{M}^{(k)}(\vecy,\tau).
\end{eqnarray}

\noindent For $k\!=\!0$, this yields $\mom{0}(\vecy)\!=\!p(\vecy,t)$ (suppressing the unneeded argument $\tau$ and
taking into account the scalar nature of $\mom{0}$). Consequently one can write $\mbf{M}^{(k)}\!=\!\bmom{k}/\mom{0}$
and one obtains
\begin{eqnarray}\label{relation_m_D}
\frac{\bmom{k}(\vecy,\tau)}{\mom{0}(\vecy)} &=& \tau\mbf{D}^{(k)}(\vecy)\!+\!O(\tau^2),\quad k=1,2.
\end{eqnarray}

\noindent The moments $\bmom{k}(\vecy,\tau)$ can directly be estimated from a given time series. In practise, this is
usually done by applying a binning approach. Estimating the moments for a number of time increments $\tau$ then allows to solve
Eq.~(\ref{relation_m_D}) for $\mbf{D}^{(k)}(\vecy)$ in a least square sense. Usually a low order polynomial in $\tau$
is used for a fit of the right hand side, as the higher order terms in above equation are known
to be powers of $\tau$ \cite{sampling02a,sampling08}. This strategy will be denoted as standard Markov analysis (SMA) in the following.

Next, a stochastic process $\vecX(t)$ is looked at that obeys the {\em second order} ODE
\begin{eqnarray}\label{second_order_ode}
\ddot \vecX &=& \vecf(\vecX,\dot\vecX)+\matg(\vecX,\dot\vecX)\,\vecxi(t),\quad\vecX\in\mathbb{R}^N.
\end{eqnarray}

\noindent Here $\vecxi(t)$ denotes Gaussian white noise again. As Eq.~(\ref{second_order_ode}) is a second order
ODE, such a process is not Markovian, i.e., the statistics of its increments do not only depend on the value of $\vecX$
but also on its derivative. In an extended phase space, however, consisting of the values of $\vecX$ and $\dot\vecX$,
the dynamic becomes Markovian. With the definitions
\begin{eqnarray}\label{def_pos_velo}
\vecY_1(t) &:=& \vecX(t),\quad \vecY_2(t)\;:=\;\dot\vecX(t),
\end{eqnarray}

\noindent Eq.~(\ref{second_order_ode}) can be written as a system of first order equations that define a Langevin
process $\vecY^t(t)\!:=\![\vecY_1^t(t),\vecY_2^t(t)]$ in $2N$ dimensions
\begin{eqnarray}\label{Langevin_Yemb}
\dot\vecY &=& \begin{bmatrix} \dot\vecY_1\\ \dot\vecY_2 \end{bmatrix}
         \;=\; \left[\begin{array}{l} \vecY_2 \\ \vecf(\vecY)+\matg(\vecY)\,\vecxi(t) \end{array}\right].
\end{eqnarray}

\noindent The Kramers--Moyal coefficients of the corresponding Fokker--Planck equation are simpler than in the
general $2N$-dimensional case, as they are given by
\begin{eqnarray}\label{kramers_moyal}
\vecDdrift(\vecy) &=& \begin{bmatrix} \vecy_2\\ \vecf(\vecy) \end{bmatrix},\quad\!
\vecDdiff(\vecy) \;=\; \begin{bmatrix} \mbf{0} &\mbf{0} \\ \mbf{0}& \matg\matg^t\!(\vecy) \end{bmatrix}\!.
\end{eqnarray}

\noindent Of cause, these coefficients can be estimated from a given time series of $\vecY(t)$ by the above mentioned
SMA. But therefor the values of $\vecX$ {\em and} $\dot\vecX$ must be given (Eq.~(\ref{def_pos_velo})). For
real word data this will not always be the case. Frequently only a series of 'positions'
$\vecY_1(t)\!\equiv\!\vecX(t)$ will be given for a second order process obeying Eq.~(\ref{second_order_ode}),
while the corresponding 'velocities' $\vecY_2(t)\!\equiv\!\dot\vecX(t)$ are missing.
It may, e.g., be hard to accurately meassure the velocities in a given experimental setup. Or it may not have been
realized in advance that $\vecX(t)$ needs to be modelled as second order process. Or it may simply
have been assumed that a highly resolved series of position values will provide sufficiently accurate information
on the velocities.

If $\dot\vecX$ is missing, these velocity values need to be estimated numerically.
This seems to be no major problem as $\vecX(t)$ is a continuously differentiable
function. Its derivative can be estimated by a discrete differencing scheme with arbitrary accuracy -- provided
the step-size of the scheme (here and in the following denoted by $\theta$) can be chosen small enough. So for
a 'sufficiently' fine sampled series of positions the estimation-errors of the velocities will become negligible.
The standard approach for an analysis, therefore, goes like this: Choose some small step-size $\theta$ and estimate
the series $\vecY_2$ using the given series $\vecY_1$. Then apply a SMA to the resulting series $\vecY$. This
strategy will be denoted as standard embedding approach (SEA) in the following.

Such an approach, however, has its flaws. For a Markov analysis, the moments of process-{\em increments}
will be looked at (see Eq.~(\ref{moments_incY})). For these quantities the effects of the estimation errors will show to
be of importance unless the step-size $\theta$ (used for velocity estimation) can be chosen much smaller than
the time increment $\tau$ (used for increment calculation). At the same time, however, $\tau$ needs to be small
compared to the characteristic time scale $T$ of the process under investigation. Otherwise the higher order terms in
Eq.~(\ref{relation_m_D}) can no longer be approximated  by a low order polynomial.
The requirement $\theta\!\ll\!\tau\!\ll\!T$ will only rarely be fulfilled in practise as it requires data with
a very high temporal resolution (compared to the characteristic time scale $T$).

Also another source of errors has to be considered for real data: Any measurement noise that afflicts the values of $\vecY_1$
will lead to an additional error in the estimation of $\vecY_2$. For a differencing scheme with step-size $\theta$,
this error will be proportional to $\theta^{-1}$, as will be seen later (assuming uncorrelated measurement noise).
So even if the measurement noise is very small, and thus negligible for $\vecY_1$
itself, it may become important in the estimation of $\vecY_2$ for small values of $\theta$.

Above considerations imply that for real data neither the values of $\vecY_1(t)$ nor that of $\vecY_2(t)$ are known
accurately. The 'noisy' values, which {\em are} at hand, will be denoted by $\vecY^*(t)$ in the following.

The aim of this paper is, to provide of a modified embedding approach (MEA) that accounts for the errors due to differencing
scheme and meassurement noise. As a by-product also a quantitative description of the errors of the SEA
will be found. However, only {\em weak} measurement noise can be accounted for. This restriction is a consequence of
the perturbative approach that will be used. The requirements on the noise will be given later, but, roughly
speaking, the noise must be negligible for the position values and its effect on the velocity increments may at most
be of the same order as the effects of the driving stochastic force $\vecxi$.

This paper is organized as follows: In Sec.~\ref{sec_moments_noisy} the observable moments $\bsmom{k}$ of the noisy time
series will be expressed in terms of moments of the {\em noisy} values $\vecY^*(t)$ and $\vecY^*(t\!+\!\tau)$ conditioned
on the {\em true} value $\vecY(t)$.
Subsequently, based on a Taylor--\Ito\ expansion, these conditional noisy values will be expressed in terms of process
parameters, measurement noise and stochastic integrals of $\vecxi$ in Sec.~\ref{sec_cond_values}.
The resulting expressions, together with an assumption on the magnitude of the measurement noise, will lead to an explicit
description of $\bsmom{k}$ in Sec.~\ref{sec_moments_Mknu} then.
This description will serve two purposes. Firstly, the effects of the reconstruction errors of a SEA
can be quantified (Sec.~\ref{sec_systematic errors}). Secondly, a MEA can be specified that allows an accurate estimation
of the Kramers--Moyal coefficients and the properties of the measurement noise (Sec.~\ref{sec_alternative_approach}).
Subsequently a numerical test case will be specified in Sec.~\ref{sec_example}, which will be used to compare the
results of SEA and MEA with and without measurement noise (Secs.~\ref{sec_comparison_nonoise} and \ref{sec_comparison_noise}).

\section{Moments of the noisy values}
\label{sec_moments_noisy}

\noindent For a series of noisy values $\vecY^*$ only the noisy counterparts $\bsmom{k}$ of the moments $\bmom{k}$ can
be estimated. In analogy to Eq.~(\ref{def_momk}) they can be defined as
\begin{eqnarray}\label{def_smomk}
\bsmom{k}(\vecy^*) &:=& \int_{\mbf{s}}(\mbf{s}\!\!-\!\vecy^*)^k  p(\vecY^*\!\!=\!\vecy^*\!;\vecY_\tau^*\!\!=\!\mbf{s})\bdif s.
\end{eqnarray}

\noindent Here and in the following, the time arguments $t$ and $\tau$ are omitted to allow for a more compact
notation. Stochastic variables implicitely refer to time $t$ now, and the shortcut $\vecY_\tau^*$ is used to denote
$\vecY^*(t\!+\!\tau)$.

Next the moments $\bsmom{k}$ need to be related to the process parameters and the properties of the measurement noise.
As outlined in Sec.~\ref{sec_intro}, the first step will be, to express the moments $\bsmom{k}$ in terms of moments of
the conditional noisy values $\vecY^*|_{\vecY\!=\vecy}$ and $\vecY_\tau^*|_{\vecY\!=\vecy}$. This can be done as follows:
First, the PDF in Eq.~(\ref{def_smomk}) is rewritten as
\begin{eqnarray}\label{def_expand_twopoint_dens}
&&\!\!\!\!\!\!\!p(\vecY^*\!\!=\!\vecy^*\!;\vecY_\tau^*\!\!=\!\mbf{s}) \;\equiv\;
\int_{\vecy}\rho(\vecy)\cr
&& \qquad\qquad\times\, p(\vecY^*\!\!=\!\vecy^*\!;\vecY_\tau^*\!\!=\!\mbf{s}\big|\vecY\!=\!\vecy) \bdif y,
\end{eqnarray}

\noindent where $\rho(\vecy)\!:=\! p(\vecY\!=\!\vecy)$ denotes the PDF of $\vecY$.
Inserting Eq.~(\ref{def_expand_twopoint_dens}) and interchanging the order of integration thus allows to
write the moments $\bsmom{k}$ in the form
\begin{eqnarray}\label{def_convlike}
\bsmom{k}(\vecy^*) &=&
\int_{\vecy}\rho(\vecy)\,\mbf{F}^{(k)}(\vecy^*\!,\vecy)\bdif y
\end{eqnarray}
\noindent with
\begin{eqnarray}
&&\!\!\!\!\!\!\!\mbf{F}^{(k)}(\vecy^*\!,\vecy) \;=\;
\int_{\mbf{s}} (\mbf{s}\!-\!\vecy^*)^k\cr
&& \qquad\qquad\times\, p(\vecY^*\!\!=\!\vecy^*\!;\vecY_\tau^*\!\!=\!\mbf{s}\big|\vecY\!=\!\vecy)\bdif s.
\end{eqnarray}

\noindent Expressing the integral in Eq.~(\ref{def_convlike}) by a moment expansion yields (using summation
convention)
\begin{eqnarray}\label{def_mstar_M}
\smom{k}_{i_1,\ldots,i_k}(\vecy^*)
&=& \sum_{\nu=0}^\infty\frac{(-1)^\nu}{\nu!}\fracpp{y^*_{j_1}}\cdots\fracpp{y^*_{j_\nu}} \cr
&& \times\, \left[\rho(\vecy^*)M^{(k,\nu)}_{i_1,\ldots,i_k,j_1,\ldots,j_\nu}\!(\vecy^*)\right],
\end{eqnarray}

\noindent where the moments are defined as
\begin{eqnarray}
\matM^{(k,\nu)}(\vecy^*) &:=& \int_{\vecz}\mbf{F}^{(k)}(\vecz,\vecy^*)\otimes(\vecz\!-\!\vecy^*)^\nu \bdif z.
\end{eqnarray}

\noindent Here $\otimes$ denotes a dyadic product. Inserting the definition of $\mbf{F}^{(k)}$ first leads to
\begin{eqnarray}
&&\!\!\!\!\!\!\!\matM^{(k,\nu)}(\vecy^*) \;=\;
\int_{\mbf{s},\vecz}(\mbf{s}\!-\!\vecz)^k\otimes(\vecz\!-\!\vecy^*)^\nu\cr
&& \qquad\quad\quad\times\,p(\vecY^*\!\!=\!\vecz ;\vecY_\tau^*\!\!=\!\mbf{s}\big|\vecY\!=\!\vecy^*)\bdif s\bdif z.
\end{eqnarray}

\noindent Using the relation $\int_af(a)p(a|b)\!=\!\left<f(A)|b\right>$ then gives

\begin{eqnarray}
\matM^{(k,\nu)}(\vecy^*) &=&
\left<(\vecY_\tau^*\!\!-\!\vecY^*)^k\!\!\otimes\!(\vecY^*\!\!-\!\vecY)^\nu\right>\!\big|_{\vecY\!=\vecy^*}.
\end{eqnarray}

\noindent The general form of the observable moments $\bsmom{k}$ therefore reads (dropping the asterisk
on the parameter $\vecy$)
\begin{eqnarray}\label{def_mstar_final}
&&\!\!\!\!\!\!\!\smom{k}_{i_1,\ldots,i_k}(\vecy) \;=\;
\sum_{\nu=0}^\infty\frac{(-1)^\nu}{\nu!}\fracpp{y_{j_1}}\cdots\fracpp{y_{j_\nu}}\cr
&& \quad\times \left\{ \rho(\vecy)\big<[\mbf{A}^k\!(\vecy)]_{i_1,\ldots,i_k}
                                      [\mbf{B}^\nu\!(\vecy)]_{j_1,\ldots,j_\nu}\big>\right\}
\end{eqnarray}

\noindent with
\begin{subequations}\label{def_A_B}
\begin{eqnarray}
\mbf{A}(\vecy) &:=& \vecY_\tau^*\big|_{\vecY\!=\vecy}\!- \vecY^*\big|_{\vecY\!=\vecy}\;,\\
\mbf{B}(\vecy) &:=& \vecY^*\big|_{\vecY\!=\vecy}\!- \vecy.
\end{eqnarray}
\end{subequations}

\noindent This is a quite general result -- no information on how $\vecY^*(t)$ and $\vecY(t)$ are related
is used so far. This will be done in the next section, where the conditional values of $\vecY^*$ and $\vecY_\tau^*$ will
be expressed explicitly.

\section{Conditional values of $\vecY^*$}
\label{sec_cond_values}

\noindent In this section we will specify the assumptions on the measurement noise together with the details of the
differencing scheme. This will allow to express the conditional values of $\vecY^*$ and $\vecY_\tau^*$ in terms of
measurement noise and conditional values of $\vecY_1$. Based on a Taylor--\Ito\ expansion, these conditional values $\vecY_1$
can then be expressed in terms of the driving stochastic force and process parameters.

To avoid confusion, time arguments will be given explicitly again in the following. However, the shortcut
$(\ldots)|_\vecy$ will be used to indicate conditioning on $\vecY(t)\!=\!\vecy$.

The given values $\vecY^*_1(t)$ are assumed to be spoilt by additive, Gaussian distributed and temporally uncorrelated
measurement noise $\vecGamma(t)$ with an expectation value of zero and covariance matrix $\matV$
\begin{subequations}\label{definition_gamma}
\begin{eqnarray}
\left<\vecGamma(t)\right> &=& \mbf{0},\\
\left<\vecGamma(t)\vecGamma^t(t')\right> &=& \delta_{t,t'} \matV,\;\;
\delta_{t,t'} :=
\left\{\begin{array}{ll}
1\,, &\; t\!=\!t' \\
0\,, &\; t\!\ne\!t'
\end{array}\right. .
\end{eqnarray}
\end{subequations}

\noindent The noise is also assumed to be independent of $\vecxi$ and $\vecY$ (implying
$\vecGamma(t)\big|_\vecy\!\equiv\vecGamma(t)$). The conditional values $\vecY^*_1\big|_\vecy$
are thus given by
\begin{eqnarray}
\vecY^*_1(t\!+\!\Delta)\big|_\vecy &:=& \vecY_1(t\!+\!\Delta)\big|_\vecy+\vecGamma(t\!+\!\Delta).
\end{eqnarray}

\noindent For the reconstruction of $\vecY_2$ a first order forward differencing scheme with a step-size of $\theta$,
applied to the observable values $\vecY^*_1$, will be used in the following. The conditional values $\vecY^*_2\big|_\vecy$
therefore are given by
\begin{eqnarray}
\vecY^*_2(t\!+\!\Delta,\theta)\big|_\vecy
 &:=& \frac{1}{\theta}\left[\vecY_1(t\!+\!\Delta\!+\!\theta)\big|_\vecy\!-\vecY_1(t\!+\!\Delta)\big|_\vecy\right]\cr
 &&    +\frac{1}{\theta}\left[\vecGamma(t\!+\!\Delta\!+\!\theta)\!-\vecGamma(t\!+\!\Delta)\right].
\end{eqnarray}

\noindent The values $\vecY_1\big|_\vecy$ at time $t\!+\!\Delta$ can be expressed by a Taylor--\Ito\ expansion
(see App.~\ref{app_taylor_ito})
\begin{eqnarray}
\vecY_1(t\!+\!\Delta)\big|_\vecy
 &=& \vecy_1+\vecy_2\,\Delta+\vecf(\vecy)\frac{\Delta^2}{2}  \cr
&&+\matg(\vecy)\mbf{I}^{t,\Delta}+\vecR^{t,\Delta}\!(\vecy).
\end{eqnarray}

\noindent Here $\mbf{I}^{t,\Delta}$ denotes a vector of stochastic integrals that only depend on the realization
of $\vecxi$ in the interval $[t,t\!+\!\Delta)$. The components of this vector are of magnitude $O(\Delta^{3/2})$ and
have an expectation value of zero. All other expansion terms are summarized in the remainder $\vecR^{t,\Delta}\!(\vecy)$
with a magnitude of $O(\Delta^2)$ and an expectation value of $O(\Delta^3)$.

In summary, above results lead to the following expressions for $\vecY^*\big|_\vecy$
\begin{subequations}\label{def_Ys}
\begin{eqnarray}
\vecY^*_1(t)\big|_\vecy &=& \vecy_1+\vecGamma(t),\\
\vecY^*_1(t\!+\!\tau)\big|_\vecy
 &=& \vecy_1+\vecy_2\,\tau+\vecf(\vecy)\frac{\tau^2}{2}
    + \matg(\vecy)\mbf{I}^{t,\tau}\cr
&&    +\vecGamma(t\!+\!\tau)+\vecR^{t,\tau}\!(\vecy),\\
\vecY^*_2(t,\theta)\big|_\vecy
 &=& \vecy_2 + \vecf(\vecy)\frac{\theta}{2}
    + \matg(\vecy)\frac{\mbf{I}^{t,\theta}}{\theta}\cr
&&    + \frac{\vecGamma(t\!+\!\theta)\!-\!\vecGamma(t)}{\theta}
    + \frac{\vecR^{t,\theta}\!(\vecy)}{\theta},\\
\vecY^*_2(t\!+\!\tau,\theta)\big|_\vecy
 &=& \vecy_2 + \vecf(\vecy)(\tau\!+\!\frac{\theta}{2})
    + \matg(\vecy)\frac{\mbf{I}^{t,\tau\!+\!\theta}\!-\!\mbf{I}^{t,\tau}}{\theta}\cr
&&    + \frac{\vecGamma(t\!+\!\tau\!+\!\theta)\!-\!\vecGamma(t\!+\!\tau)}{\theta}\cr
&&    + \frac{\vecR^{t,\tau\!+\!\theta}\!(\vecy)\!-\!\vecR^{t,\tau}\!(\vecy)}{\theta}.
\end{eqnarray}
\end{subequations}


\section{Moments $\matM^{(k,\nu)}$}
\label{sec_moments_Mknu}

\noindent Now the moments $\matM^{(k,\nu)}$ can be attacked. For a calculation of $\matM^{(k,\nu)}$ explicit
expressions for the vectors $\mbf{A}$ and $\mbf{B}$, as defined in Eq.~(\ref{def_A_B}), are needed.
Using the results from the previous section (Eq.~(\ref{def_Ys})) one finds 
\begin{widetext}
\begin{subequations}\label{def_A_B_explicit}
\begin{eqnarray}
\mbf{A}(\vecy,\tau,\theta)
 &=&
\begin{bmatrix}
\vecy_2\,\tau + \matg(\vecy)\mbf{I}^{t,\tau}
    + \vecGamma(t\!+\!\tau) - \vecGamma(t)+\vecf(\vecy)\frac{\tau^2}{2} + \vecR^{t,\tau}\!(\vecy)\\[.3em]
\vecf(\vecy)\tau
    + \matg(\vecy)\frac{\mbf{I}^{t,\tau\!+\!\theta}-\mbf{I}^{t,\tau}-\mbf{I}^{t,\theta}}{\theta}
    + \frac{\vecGamma(t\!+\!\tau\!+\!\theta)-\vecGamma(t\!+\!\tau)-\vecGamma(t\!+\!\theta)+\vecGamma(t)}{\theta}
    + \frac{\vecR^{t,\tau\!+\!\theta}\!(\vecy)-\vecR^{t,\tau}\!(\vecy)-\vecR^{t,\theta}\!(\vecy)}{\theta}
\end{bmatrix},\\[.3em]
\mbf{B}(\vecy,\tau,\theta)
 &=&
\begin{bmatrix}
\vecGamma(t)\\[.3em]
\vecf(\vecy)\frac{\theta}{2}
    + \matg(\vecy)\frac{\mbf{I}^{t,\theta}}{\theta}
    + \frac{\vecGamma(t\!+\!\theta)-\vecGamma(t)}{\theta}
    + \frac{\vecR^{t,\theta}\!(\vecy)}{\theta}
\end{bmatrix}.
\end{eqnarray}
\end{subequations}

\end{widetext}

\noindent These expressions contain infinitely many terms, summarized in the remainders $\mbf{R}$. To allow for a
series truncation, a small parameter $\eps$ is introduced in the following to express the magnitude of terms
(it is tacitly assumed here that the problem is described in dimensionless form with $\vecf$ an $\matg$ being of order $O(1)$).
It will be assumed that $\tau$ and $\theta$ are of the same order of magnitude as $\eps$ and that the measurement noise
$\Gamma_i$ is of the same order as $\eps^{3/2}$
\begin{eqnarray}
\tau \overset{!}{=} O(\eps),\quad \theta \overset{!}{=} O(\eps),\quad V_{ij} \overset{!}{=} O(\eps^3).
\end{eqnarray}

\noindent In a strict sense, the use of the Landau symbols here is not appropriate, because there is no functional
relation between $\eps$ and, e.g., $\tau$. Above notation is rather used to express the assumptions that, firstly, $\tau$,
$\eps$ and $\Gamma_i$ are small quantities, which allows to sort powers by magnitude (like e.g. $\tau^2\!\ll\!\tau$).
Secondly, it is assumed that $\tau$, $\theta$ and $|\Gamma_i|^{2/3}$ are of 'compareable size', where compareable size
means that, when resticting to small exponents, also powers of different quantities can be sorted by size (like e.g.
$\tau^3\!\ll\!\theta^2$ or $|V_{ij}|\!\ll\!\tau^2$). This will be sufficient for appropriate low order approximations.

With this assumptions the lowest order terms in $\mbf{A}$ and $\mbf{B}$ are of order $O(\eps^{1/2})$.
The magnitude of a moment $\matM^{(k,\nu)}$, therefore, is given by (omitting arguments)
\begin{eqnarray}
\matM^{(k,\nu)} &=& \left<\mbf{A}^k\otimes\mbf{B}^\nu\right> \;=\; O(\eps^{(k\!+\!\nu)/2}).
\end{eqnarray}

\noindent For a first order description of the moments $\bsmom{k}$ thus only moments
$\matM^{(k,\nu)}$ with $k\!+\!\nu\le 2$ need to be taken into account. Using Eq.~(\ref{def_var_Ii0}) and the properties
of $\vecGamma$, one finds
\begin{subequations}
\begin{eqnarray}
\matM^{(0,0)} &=& 1,\\[0.5em]
\matM^{(0,1)} &=& \begin{bmatrix} \mbf{0} \\ \frac{1}{2}\theta\vecf \end{bmatrix}\!+\!O(\eps^2),\\
\matM^{(0,2)} &=&
\begin{bmatrix}
  \mbf{0} & \mbf{0} \\
  \mbf{0} & \frac{1}{3}\theta\matg\matg^t\!+\!2\frac{\matV}{\theta^2}
\end{bmatrix}\!+\!O(\eps^2),
\end{eqnarray}
\end{subequations}
\begin{subequations}
\begin{eqnarray}
\matM^{(1,0)} &=& \begin{bmatrix} \tau\vecy_2 \\ \tau\vecf \end{bmatrix}\!+\!O(\eps^2),\\
\matM^{(1,1)} &=&
\begin{bmatrix}
  \mbf{0} & \mbf{0} \\
  \mbf{0} &  \frac{\tau\!-\!\psi}{2}\matg\matg^t\!-\!(2\!+\!\delta_{\tau,\theta})\frac{\matV}{\theta^2}
\end{bmatrix}\!+\!O(\eps^2),
\end{eqnarray}
\end{subequations}
\begin{eqnarray}
\matM^{(2,0)} &=&
\begin{bmatrix}
  \mbf{0} & \mbf{0} \\
  \mbf{0} & \psi\,\matg\matg^t\!+\!2(2\!+\!\delta_{\tau,\theta})\frac{\matV}{\theta^2}
\end{bmatrix}\!+\!O(\eps^2),
\end{eqnarray}

\noindent with
\begin{eqnarray}\label{def_psi}
\psi &:=& \left\{
\begin{array}{lcl}
\tau^2/\theta-\frac{1}{3}\tau^3/\theta^2 &,&\tau<\theta\\[0.3em]
\tau-\frac{1}{3}\theta   &,&\tau\ge\theta
\end{array}\right. .
\end{eqnarray}

\noindent Inserting these expressions into Eq.~(\ref{def_mstar_M}), finally, yields a first order
description of the moments $\bsmom{k}$ in terms of $\rho$, $\vecf$, $\matg$ and $\matV$.
It turns out that derivatives with respect to components of $\vecy_1$ do not appear in the terms up to order $O(\eps)$
-- so for a first order description only the derivatives with respect to the components of $\vecy_2$ need to be considered.
It also turns out that only the upper half of the vector $\bsmom{1}$ and the upper quarter of the matrix $\bsmom{2}$ need to
be looked at (those components that correspond to moments of the increments of $\vecY^*_2$). To take (syntactical)
advantage of this reduction in dimensionality the notations
\begin{subequations}
\begin{eqnarray}
\hat\p_i &:=& \fracpp{y_{N\!+\!i}},\qquad     \hmom{0} \;:=\; \smom{0},\\
\hmom{1}_i &:=& \smom{1}_{N\!+\!i},\qquad\;   \hmom{2}_{ij} \;:=\; \smom{2}_{N\!+\!i,N\!+\!j},
\end{eqnarray}
\end{subequations}

\noindent are introduced, where $i$ and $j$ are in the range $1,\ldots,N$.
The relevant equations can now be written compactly as
\begin{subequations}\label{relation_mstar_D}
\begin{eqnarray}
\label{relation_mstar0_D}
\hmom{0}(\vecy,\theta) &=& \rho
            -\frac{\theta}{2}\,\hat\p_i [\rho f_i]
            +\frac{\theta}{6}\,\hat\p_i\hat\p_j [\rho\,(\matg\matg^t)_{ij}]\cr
&&            +\frac{V_{ij}}{\theta^2}\,\hat\p_i\hat\p_j\rho +O(\eps^2),\\[0.5em]
\hmom{1}_i(\vecy,\tau,\theta) &=& \tau\rho f_i
              -\frac{1}{2}(\tau\!-\!\psi)\,\hat\p_j [\rho\,(\matg\matg^t)_{ij}]\cr
&&              +(2\!+\!\delta_{\tau,\theta})\frac{V_{ij}}{\theta^2}\,\hat\p_j\rho +O(\eps^2),\\[0.5em]
\hmom{2}_{ij}(\vecy,\tau,\theta) &=&
            \psi\rho\,(\matg\matg^t)_{ij}\cr
&&            + 2(2\!+\!\delta_{\tau,\theta})\frac{V_{ij}}{\theta^2}\,\rho+O(\eps^2).
\end{eqnarray}
\end{subequations}

\noindent These equations directly relate the unknown quantities $\rho$, $\vecf$, $\matg\matg^t$ and $\matV$ and the
observable quantities $\bhmom{k}$. The function argument
of $\rho$, $\vecf$ and $\matg$ is given by $\vecy$. The function $\psi$ depends on $\tau$ and $\theta$ and has
a piecewise definition only, Eq.~(\ref{def_psi}).

\section{Systematic errors of the standard embedding approach (SEA)}
\label{sec_systematic errors}

\noindent Next the SEA will be analyzed, using the final result of the previous section (Eq.~(\ref{relation_mstar_D})).
Only the case without measurement noise, i.e. $\matV\equiv \mbf{0}$, will be looked at. This will show
the 'pure' effects of the reconstruction errors caused by the numerical estimation of $\vecY_2$.

For a time series, where $\vecY_2$ has been reconstructed by a first order forward differencing scheme with stepsize
$\theta$, the observable moments $\bhmom{k}$ are described by
Eq.~(\ref{relation_mstar_D}). Ignoring this result and attempting a Markov analysis as outlined in Sec.~\ref{sec_intro}
will put the focus on the terms $\bhmom{k}(\vecy,\tau,\theta)/(\tau\hmom{0}(\vecy,\theta))$. According to
Eq.~(\ref{relation_m_D}), these terms should be finite-increment estimates of $\vecf$ and $\matg\matg^t$ ($k\!=\!1$ resp.
$k\!=\!2$). In fact, however, the terms evaluate to
\begin{subequations}\label{estim1_f_g}
\begin{eqnarray}
\label{estim1_f}
\frac{\hmom{1}_i(\vecy,\tau,\theta)}{\tau\hmom{0}(\vecy,\theta)}
  &=& f_i -\frac{1\!-\!\psi(\tau,\theta)/\tau}{2\hmom{0}}\,\hat\p_j \left[\hmom{0}\,(\matg\matg^t)_{ij}\right]\cr
&& +O(\eps)\\
\label{estim1_g}
\frac{\hmom{2}_{ij}(\vecy,\tau,\theta)}{\tau\hmom{0}(\vecy,\theta)}
  &=& \frac{\psi(\tau,\theta)}{\tau}(\matg\matg^t)_{ij} +O(\eps).
\end{eqnarray}
\end{subequations}

\noindent Trying to extrapolate these estimates to $\tau\!=\!0$ then becomes problematic. Instead of being approximately
constant, as expected from Eq.~(\ref{relation_m_D}), the values will show
non-linear behaviour caused by the function $\psi/\tau$.
For fixed $\theta$ this function starts linear with a value of zero at $\tau/\theta\!=\!0$, passes through $2/3$ at
$\tau/\theta\!=\!1$ and approaches a value of one for $\tau/\theta\!\to\!\infty$.
Simply fitting a low order polynomial to all estimates up to some maximum increment $\tau_\text{max}$ will thus,
in general, under-estimate $\matg\matg^t$ (because of $|\psi/\tau|<1$). An error of compareable size (although with
arbitrary sign) will occure when estimating $\vecf$.

In principle, however, the estimates for large $\tau$, i.e. where $\psi/\tau\approx 1$, could be used for a fit.
On the other hand also the influence of higher order terms becomes stronger for large increments. Unless a time series is
sampled with a very small timestep, such an approach will also fail to provide accurate estimates for $f_i$ and
$(\matg\matg^t)_{ij}$.

\section{Modified embedding approach (MEA)}
\label{sec_alternative_approach}

\noindent Based on Eq.~(\ref{relation_mstar_D}), we now will propose a modified approach that takes into account the
effects of the differencing scheme as well as the effects of measurement noise. An important point in this approach
will be to keep the ratio of $\tau$ and $\theta$ fix. This provides an easy way to avoid problems caused by the
non-linear term $\psi(\tau,\theta)$. In the following $\theta\!\equiv\!\tau$ is chosen. Equation~(\ref{relation_mstar_D})
then reads
\begin{subequations}\label{relation2_hatm_D}
\begin{eqnarray}
\label{relation2_hatm0_D}
\hmom{0}(\vecy,\tau) &=& \rho
            -\frac{\tau}{2}\,\hat\p_i [\rho f_i]
            +\frac{\tau}{6}\,\hat\p_i\hat\p_j [\rho\,(\matg\matg^t)_{ij}]\cr
&&            +\frac{V_{ij}}{\tau^2}\,\hat\p_i\hat\p_j\rho +O(\eps^2), \\
\label{relation2_hatm1_D}
\hmom{1}_i(\vecy,\tau,\tau) &=& \tau\rho f_i
              -\frac{\tau}{6}\,\hat\p_j [\rho\,(\matg\matg^t)_{ij}]\cr
&&              +3\frac{V_{ij}}{\tau^2}\,\hat\p_j\rho +O(\eps^2), \\
\label{relation2_hatm2_D}
\hmom{2}_{ij}(\vecy,\tau,\tau) &=&
            \tau\frac{2}{3}\rho\,(\matg\matg^t)_{ij}\cr
&&            + 6\frac{V_{ij}}{\tau^2}\,\rho +O(\eps^2).
\end{eqnarray}
\end{subequations}

\noindent A fixed ratio of $\tau$ and $\theta$ also leads to a simpler form of the higher order terms
(see App.~\ref{app_higher_order}). Each term of order $O(\eps^n)$ on the right hand side of Eq.~(\ref{relation2_hatm_D})
has the form
\begin{eqnarray}\label{def_high_order}
Q^{(n)} &=& c(\vecy)\,\tau^a(\eps^3/\tau^2)^b
\end{eqnarray}

\noindent with
\begin{eqnarray}
0\le a\le n,\quad b=n-a,\quad c=O(1).
\end{eqnarray}

\noindent Here the symbol $Q^{(n)}$ is used to denote such a term and $\eps^3$ accounts for the assumption on the magnitude
of $\matV$. The functional form of $Q^{(n)}$ (with respect to $\tau$) can thus be described by a function-base ${\cal B}^{(n)}$
that consists of $n\!+\!1$ functions $\tau^{a-2b}$. As noted in App.~\ref{app_higher_order}, this implies
${\cal B}^{(n)}\!\subset{\cal B}^{(n\!+\!3)}$ and thus puts a limit
on the accuracy that can be achieved in least square fits of $\bhmom{k}$. E.g. it is not possible to distinguish a first
order term $c\tau$ and a fourth order term $c'\tau^3(\eps^3/\tau^2)$ by their functional form.

In the following Eq.~(\ref{relation2_hatm0_D}) will be used in the form $\hmom{0}\!=\!\rho+O(\eps)$, i.e. the explicit
results for the first order terms will not be used. This avoids the need to numerically calculate the derivatives that
appear within these terms. Next, Eqs.~(\ref{relation2_hatm1_D}) and (\ref{relation2_hatm2_D}) are divided by $\rho$.
Replacing $\rho$ by $\hmom{0}$ in the resulting left hand sides will only result in additionally terms of order
$O(\eps^2)$ and higher for the right hand sides. One finds (omitting function arguments again)
\begin{subequations}\label{relation3_hatm_D}
\begin{eqnarray}
\label{relation3_hatm1_D}
\frac{\hmom{1}_i}{\hmom{0}} &=& \tau\tilde f_i+\frac{3V_{ij}\hat\p_j\rho}{\tau^2\rho}    +O(\eps^2),\\
\label{relation3_hatm2_D}
\frac{\hmom{2}_{ij}}{\hmom{0}} &=& \tau\frac{2}{3}\,(\matg\matg^t)_{ij}+\frac{6V_{ij}}{\tau^2}+O(\eps^2),
\end{eqnarray}
\end{subequations}

\noindent with the shortcut
\begin{eqnarray}\label{correction_f}
\tilde f_i &:=& f_i-\frac{\hat\p_j [\rho\,(\matg\matg^t)_{ij}]}{6\rho}.
\end{eqnarray}

\noindent The term $3V_{ij}(\hat\p_j\rho)/(\tau^2\rho)$ in Eq.~(\ref{relation3_hatm1_D}) will now be expressed as
$\tau^{-2}c_i$, where $c_i$ is a unknown constant (of order $O(\eps^3)$). Finally, it will be assumed that
$\matV$ is known. This assumption is not mandatory -- $\matV$ could be estimated using Eq.~(\ref{relation3_hatm2_D}) --
but this quantity can be estimated more easily in advance by, e.g., analyzing the auto-covariance
of $\vecY^*_1$. The final set of equations now reads
\begin{subequations}\label{relation4_hatm_D}
\begin{eqnarray}
\label{relation4_hatm0_D}
\hmom{0} &=& \rho+O(\eps), \\
\label{relation4_hatm1_D}
\frac{\hmom{1}_i}{\hmom{0}} &=& \tau\tilde f_i+\frac{1}{\tau^2}c_i    +O(\eps^2),\\
\label{relation4_hatm2_D}
\frac{\hmom{2}_{ij}}{\hmom{0}}-\frac{6V_{ij}}{\tau^2} &=& \tau\frac{2}{3}(\matg\matg^t)_{ij}+O(\eps^2).
\end{eqnarray}
\end{subequations}

\noindent The terms on the left hand sides can be estimated for different values of $\tau$ from a given time series.
Choosing appropriate sets of regression functions thus allows to estimate $\rho$, $\tilde\vecf$ and $\matg\matg^t$ by a
linear regression analysis. Once these quantities have been estimated, Eq.~(\ref{correction_f}) can be
used to finally calculate $\vecf$ (the derivative that appears in Eq.~(\ref{correction_f}) can, e.g., be calculated
using a density-weighted local polynomial fit of $\rho\matg\matg^t$).

The functional form of the higher order terms can be shown to still obey Eq.~(\ref{def_high_order}).
Therefore $\{1,\tau,\tau^{-2}\}$, $\{\tau,\tau^{-2}\}$ and $\{\tau\}$ are appropriate function sets for
Eq.~(\ref{relation4_hatm0_D}), (\ref{relation4_hatm1_D}) and (\ref{relation4_hatm2_D}) if terms up to
order $O(\eps)$ shall be taken into account. To also take into account second order terms, the functions
$\{\tau^2\!,\tau^{-1}\!,\tau^{-4}\}$ must be added to the sets. In principle, also third order terms can be accounted
for in Eq.~(\ref{relation4_hatm1_D}) and (\ref{relation4_hatm2_D}) by also adding the functions
$\{\tau^3\!,1,\tau^{-3}\!,\tau^{-6}\}$. In practise, however, a large number of regression functions and also large
negative exponents lead to numerical problems. As a compromise, the terms can partially be accounted for. In the
numerical example given later, e.g., only $\tau^3$ is used as regression function for third order terms.

\section{Numerical test case}
\label{sec_example}

\noindent To check the analytical results and to compare the different embedding approaches, a numerical
example is investigated now. As test case a scalar process $X(t$) is chosen that obeys the second order ODE
\begin{eqnarray}\label{ex_def_X}
\ddot X &=& f(X,\dot X)+g(X,\dot X)\xi(t),
\end{eqnarray}

\noindent where $f$ and $g$ are defined as
\begin{eqnarray}\label{def_exp_f_g}
f(X,\dot X) &:=& -X-3\dot X,\quad g(X,\dot X) \;:=\; 1.
\end{eqnarray}

\noindent Again $\xi(t)$ denotes Gaussian white noise with $\left<\xi(t)\xi(t')\right>\!=\!\delta(t\!-\!t')$.
Above ODE can be rewritten as a system of first order ODEs for a $2D$ process $\vecY(t)$, the components of which
are given by position $Y_1\!\equiv\!X$ and velocity $Y_2\!\equiv\!\dot X$ of the $1D$ process $X(t)$
\begin{subequations}
\begin{eqnarray}\label{ex_def_Y}
\dot Y_1 &=& Y_2,\\
\dot Y_2 &=& -Y_1-3Y_2+\xi.
\end{eqnarray}
\end{subequations}

\noindent These equations describe an Ornstein--Uhlenbeck process in two dimensions and can be solved analytically.
The characteristic time scales of the auto-covariance of $\vecY$ are found to be $(3\!+\!\sqrt{5})/2\!\approx\!2.618$
and $(3\!-\!\sqrt{5})/2\!\approx\!0.382$. The values of $\vecY$ are Gaussian distributed and have a variance of
$\left<\vecY\vecY^t\right>\!=\!\mbf{Id}/6$.

\begin{figure}[h]
\begin{center}
\includegraphics*[width=7.5cm,angle=0]{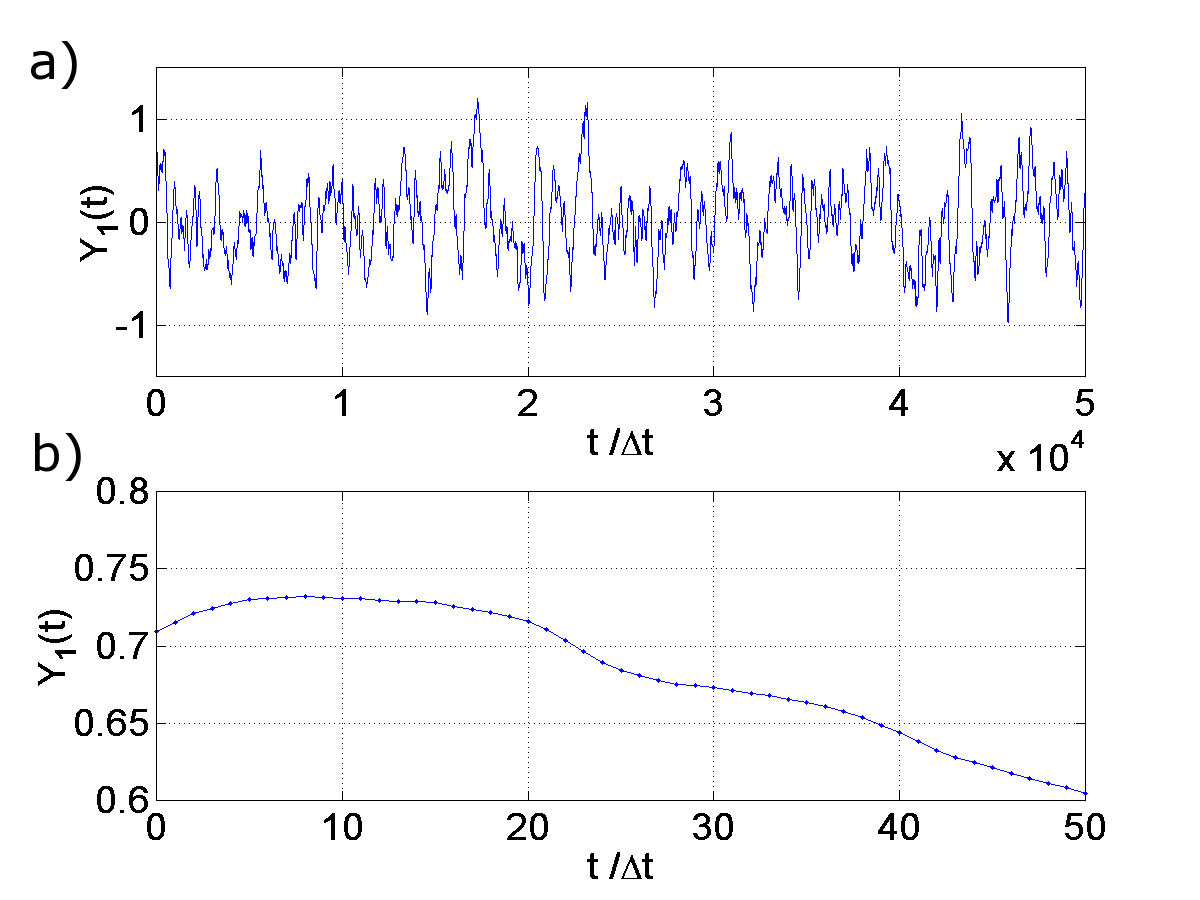}\vspace{-2em}
\end{center}
\caption{ Excerpt of the generated series of position values $Y_1$ ({\bf a}). A zoomed-in view ({\bf b})
shows that the signal in fact is smooth and thus allows to numerically estimate its derivative if the sampling time step
$\Delta t$ is sufficiently small.}
\label{fig1}
\end{figure}

\begin{figure}[h]
\begin{center}
\includegraphics*[width=7.5cm,angle=0]{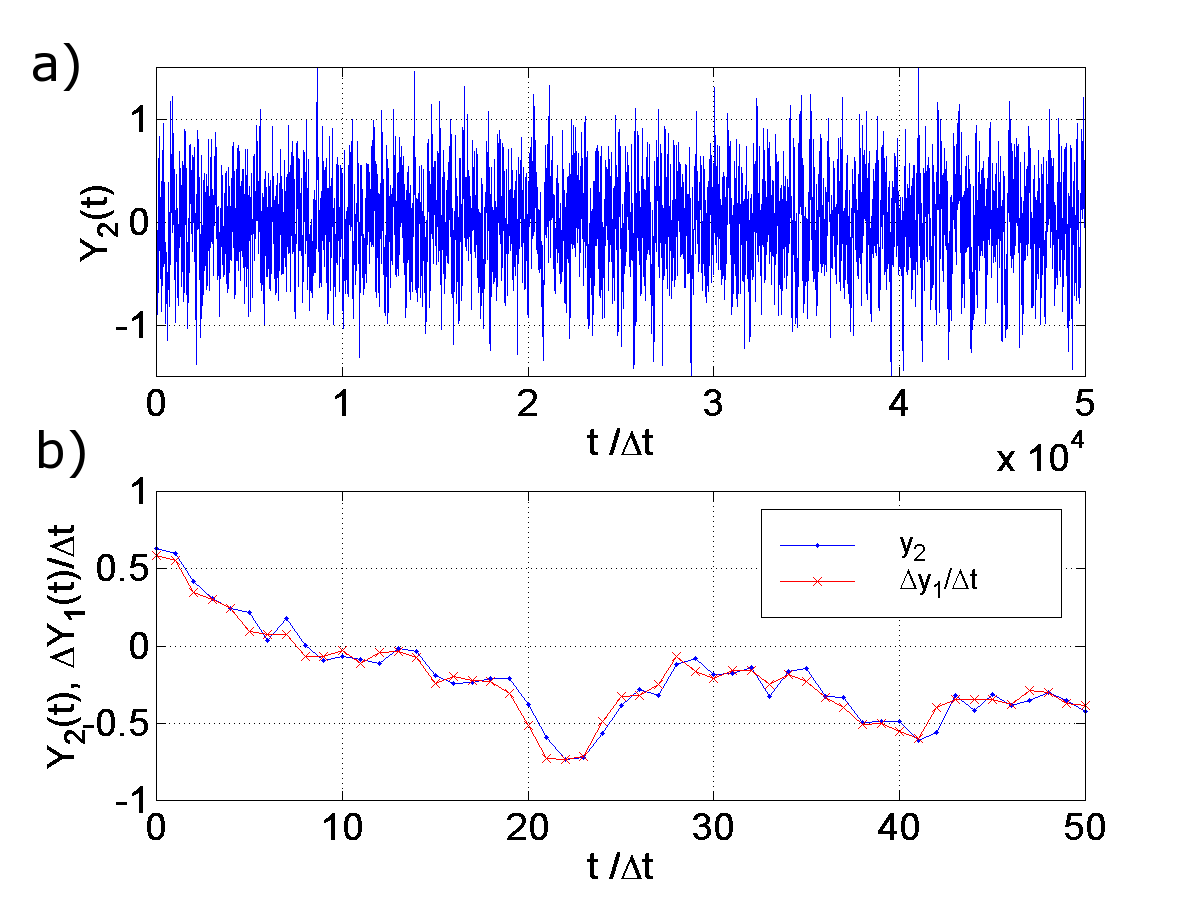}\vspace{-2em}
\end{center}
\caption{ Excerpt of the generated series of velocity values $Y_2$ ({\bf a}). In the zoomed-in view ({\bf b}) additionally
the numerically estimated derivative of $Y_1$ is shown. Allthough the values of both series (true and estimated) quite
accurately match, there are notable differences for the small scale increments.}
\label{fig2}
\end{figure}

For this process a time series of $\vecY$, consisting of $10^7$ values, sampled with a time increment $\Delta t\!=\!0.01$,
is generated. Excerpts of the resulting series for $Y_1$ and $Y_2$ are shown in Figs.~(\ref{fig1}) and (\ref{fig2}).
Here also a basic problem of the SEA can be seen, which was noted in Sec.~(\ref{sec_intro}) and quantified
in Sec.~(\ref{sec_systematic errors}): Even if a series is sampled sufficiently fine to allow an 'accurate' estimation of
$Y_2$ by a numerical differencing scheme, the velocity {\em increments} (for time increment $\tau$) will still show notable
errors for small $\tau$. This error depends on the ratio $\theta/\tau$ (here $\theta\!=\!\Delta t$) and its effects can be
quantified by the function $\psi$ in Eq.~(\ref{estim1_f_g}).

\begin{figure}[h]
\begin{center}
\includegraphics*[width=7.5cm,angle=0]{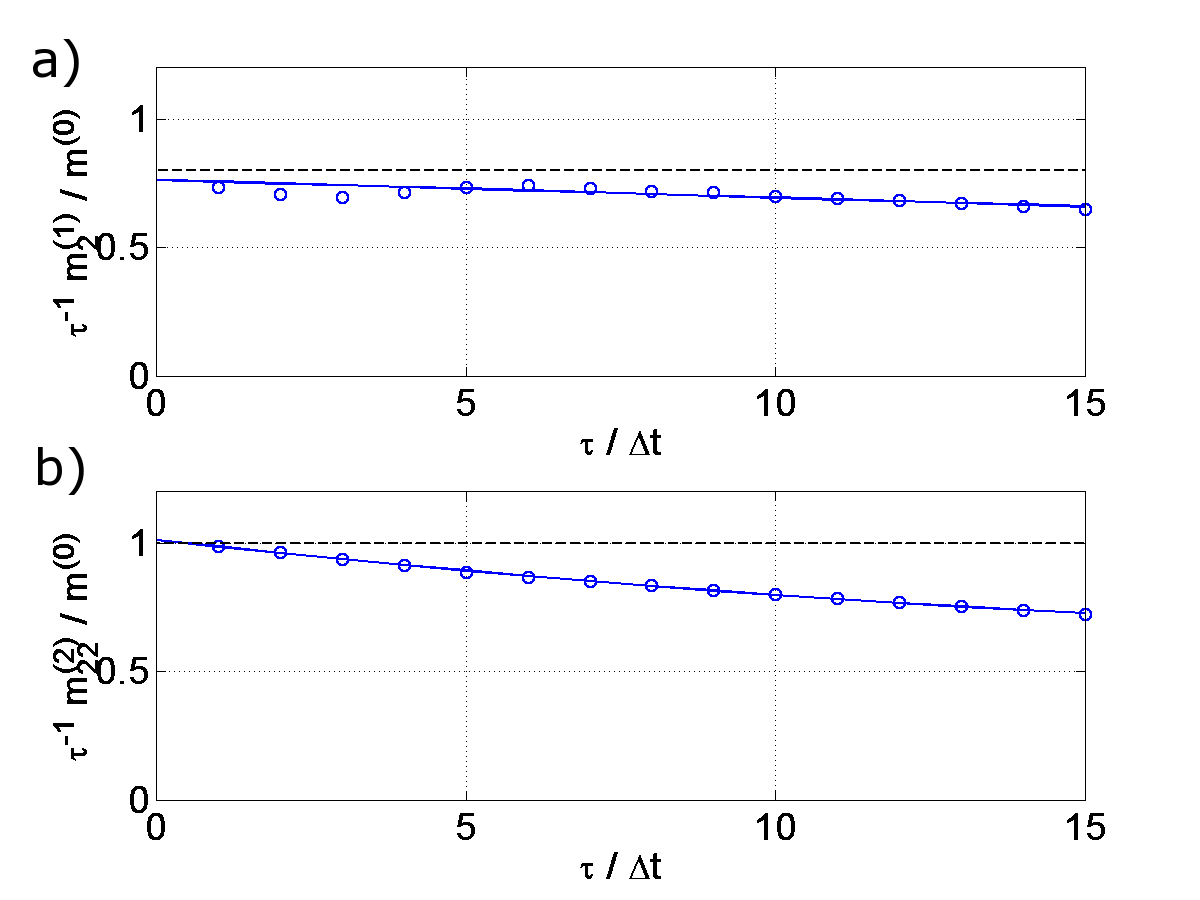}\vspace{-2em}
\end{center}
\caption{ First ({\bf a}) and second moment ({\bf b}) of the conditional velocity increments (obtained by
a SMA). The estimated values (circles) are scaled by $\tau^{-1}$. The corresponding fits are
shown as solid curves. Estimates are taken at $(y_1,y_2)\!=\!(-0.1,-0.2333)$. Here $f$ and $g^2$ have values of
$0.8$ and $1.0$ respectively (dashed lines).}
\label{fig3}
\end{figure}

To obtain a baseline for the accuracy that can be achieved with the given data, a SMA is applied to the
true $2D$ series $\vecY$ first.
Here and for subsequent analyses a binning approach is used, where the region
$[-1,1]\times[-1,1]$ of the $(y_1,y_2)$-plane is covered by $30\times30$ bins. For one of these bins the estimated
moments of the conditional velocity increments are shown in Fig.~(\ref{fig3}).

Actually the moments in Fig.~(\ref{fig3}) are scaled by $\tau^{-1}$, as is usually done for their visual
presentation. This allows to interprete the estimation of $f$ and $g^2$ as 'extrapolating the scaled moments to $\tau\!=\!0$'.
Later on, however, when measurement noise enters the scene, a more general interpretation will be needed, where $f$ and $g^2$
are found by a linear regression strategy. Of cause this interpretation is also valid in the given setup. The values of
$f$ and $g^2$ are given by the coefficients of the linear part (in $\tau$) of the conditional moments $\mom{1}_2/\mom{0}$
and  $\mom{2}_{22}/\mom{0}$ respectively (see Eq.~(\ref{relation_m_D})).

\begin{figure}[h]
\begin{center}
\includegraphics*[width=8.6cm,angle=0]{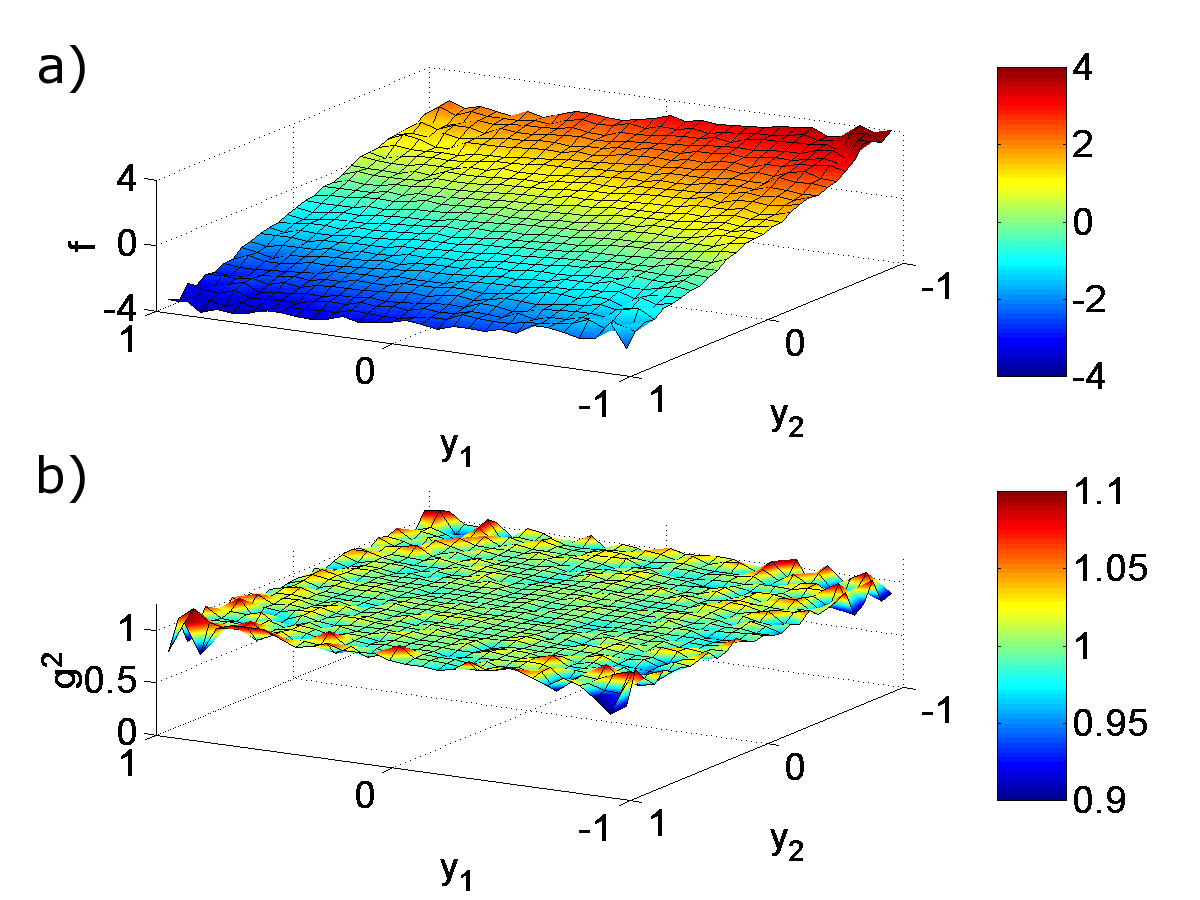}\vspace{-2em}
\end{center}
\caption{ Estimates for $f$ and $g^2$ ({\bf a}, {\bf b}), obtained by a SMA.}
\label{fig4}
\end{figure}

\begin{figure}[h]
\begin{center}
\includegraphics*[width=8.6cm,angle=0]{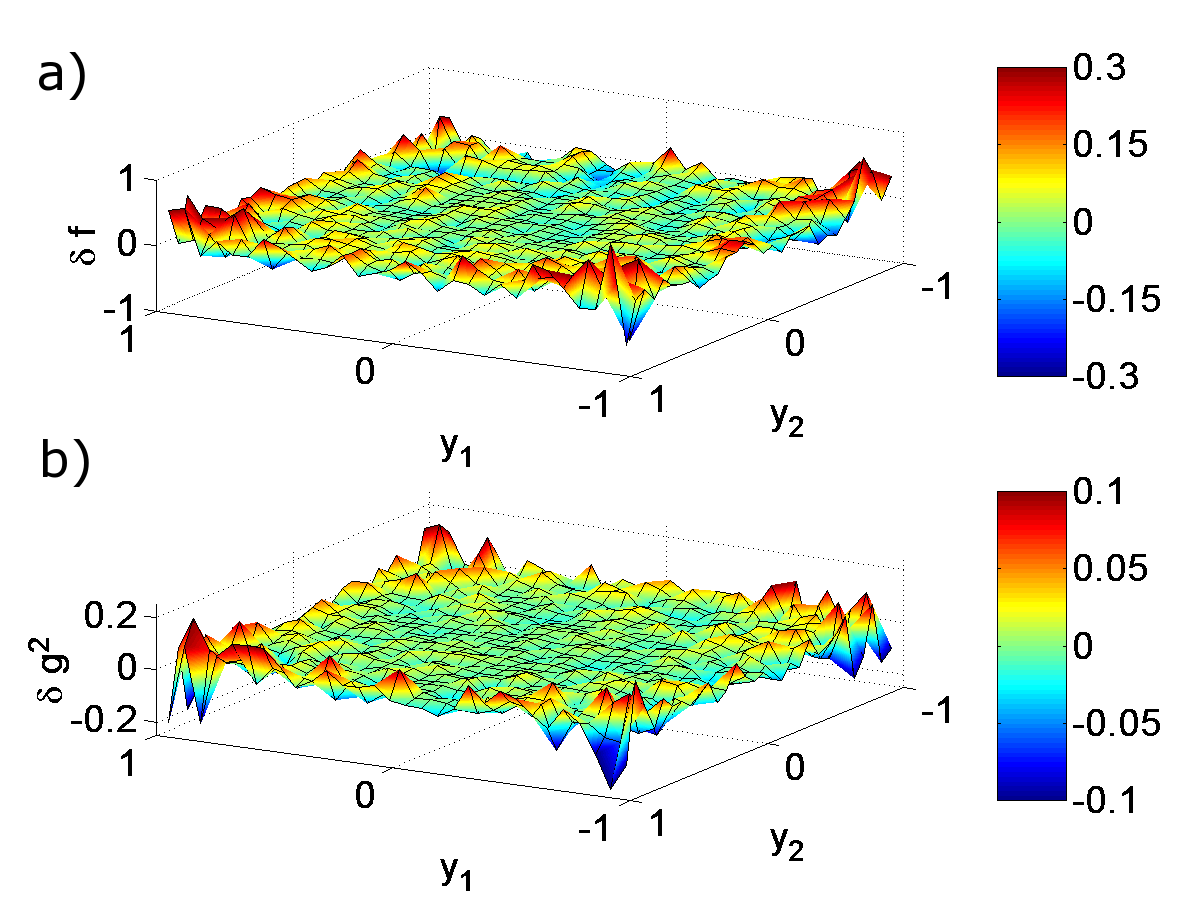}\vspace{-2em}
\end{center}
\caption{ Absolute errors of the estimates for $f$ and $g^2$ ({\bf a}, {\bf b}), obtained by a SMA.}
\label{fig5}
\end{figure}

In the following the regression functions $\{\tau,\tau^2\}$ and $\{\tau,\tau^2,\tau^3\}$ are used to fit the estimated
first and second conditional moments (this corresponds to fitting a linear function to the values in Fig.(\ref{fig3}a) and
a quadratic function to those in Fig.(\ref{fig3}b)). The maximum increment that is used for these fits is chosen
as $\tau_\text{max}\!=\! 15\Delta t$. The resulting estimates for $f$ and $g^2$ are shown in Fig.~(\ref{fig4}). In
Fig.~(\ref{fig5}) the absolute errors $\delta f$ and $\delta g^2$ of these estimates are shown. In regions with low
density (as noted above, the PDF of $\vecY$ is a symmetric Gaussian with a standard deviation of $\approx 0.408$) fluctuations
become larger but there is no obvious bias of the results.

\section{Embedding approaches without measurement noise}
\label{sec_comparison_nonoise}

\noindent Next the results of the different embedding approaches are looked at. First a SEA is used to perform an
analysis solely based on the $1D$ series of positions $Y_1$. The corresponding velocity values $Y_2$ are estimated by
a first order forward differencing scheme with a step size of $\theta\!=\!\Delta t$ and the resulting $2D$ series then
is analyzed by a SMA.

\begin{figure}[h]
\begin{center}
\includegraphics*[width=7.5cm,angle=0]{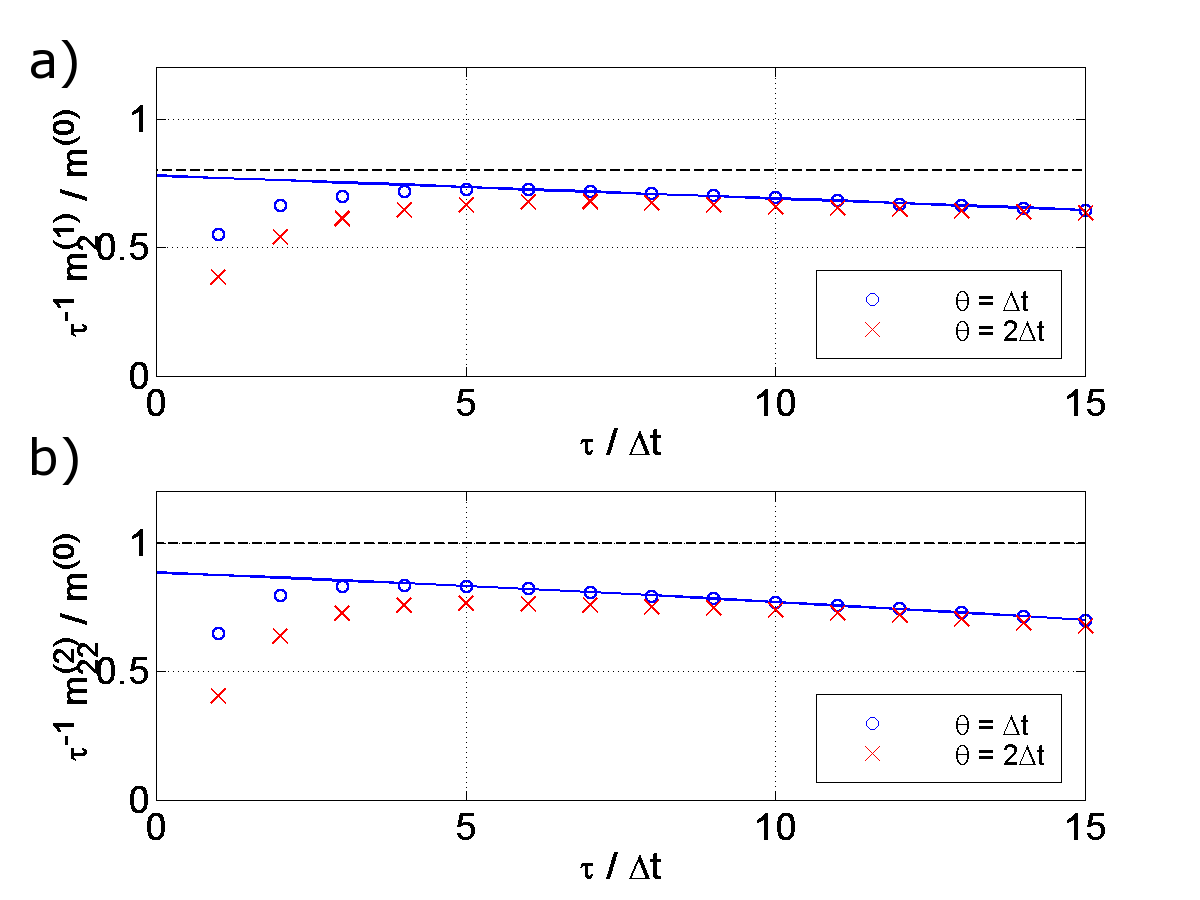}\vspace{-2em}
\end{center}
\caption{ First ({\bf a}) and second moment ({\bf b}) of the conditional velocity increments (obtained by
a SEA with $\theta\!=\!\Delta t$). The estimated values (circles) are scaled by $\tau^{-1}$. The
corresponding fits are shown as solid curves. Additionally, the estimates obtained by a SEA with
$\theta\!=\!2\Delta t$ are shown (crosses). Estimates are taken at $(y_1,y_2)\!=\!(-0.1,-0.2333)$. Here $f$ and
$g^2$ have values of $0.8$ and $1.0$ respectively (dashed lines).}
\label{fig6}
\end{figure}

\begin{figure}[h]
\begin{center}
\includegraphics*[width=8.6cm,angle=0]{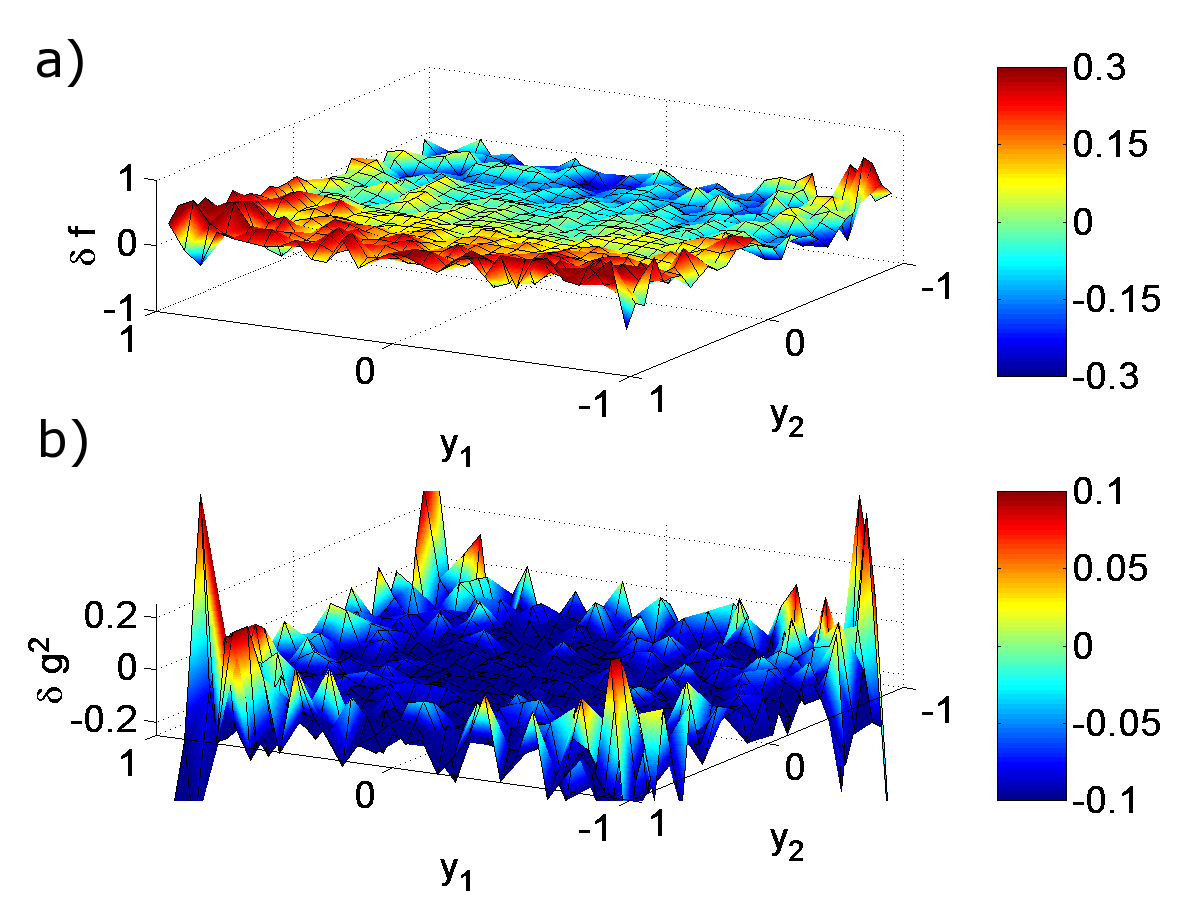}\vspace{-2em}
\end{center}
\caption{ Absolute errors of the estimates for $f$ and $g^2$ ({\bf a}, {\bf b}), obtained by a SEA.}
\label{fig7}
\end{figure}

As is shown in Fig.~(\ref{fig6}), the estimated moments of the conditional velocity increments behave quite different,
compared to those obtained from the true $2D$ series (shown in Fig.~(\ref{fig3})). As
expected from Eq.~(\ref{estim1_f_g}), the moments show strongly nonlinear behaviour for small increments $\tau$. For
an estimation of $f$ and $g^2$, therefore, only increments with $5 \le\tau/\Delta t \le 15$ are used. Least square
fits are performed using the same sets of regression functions as in the previous section. The absolute errors $\delta f$
and $\delta g^2$ of the resulting estimates are shown in Fig.~(\ref{fig7}). Of cause, the fluctuations become larger
now as fewer increments are used for the fits. But, more importantly, it is obvious that $g^2$ is systematically
under-estimated. And also the estimates for $f$ clearly show a significant bias that is approximately linear in $y_2$.

\begin{figure}[h]
\begin{center}
\includegraphics*[width=8.6cm,angle=0]{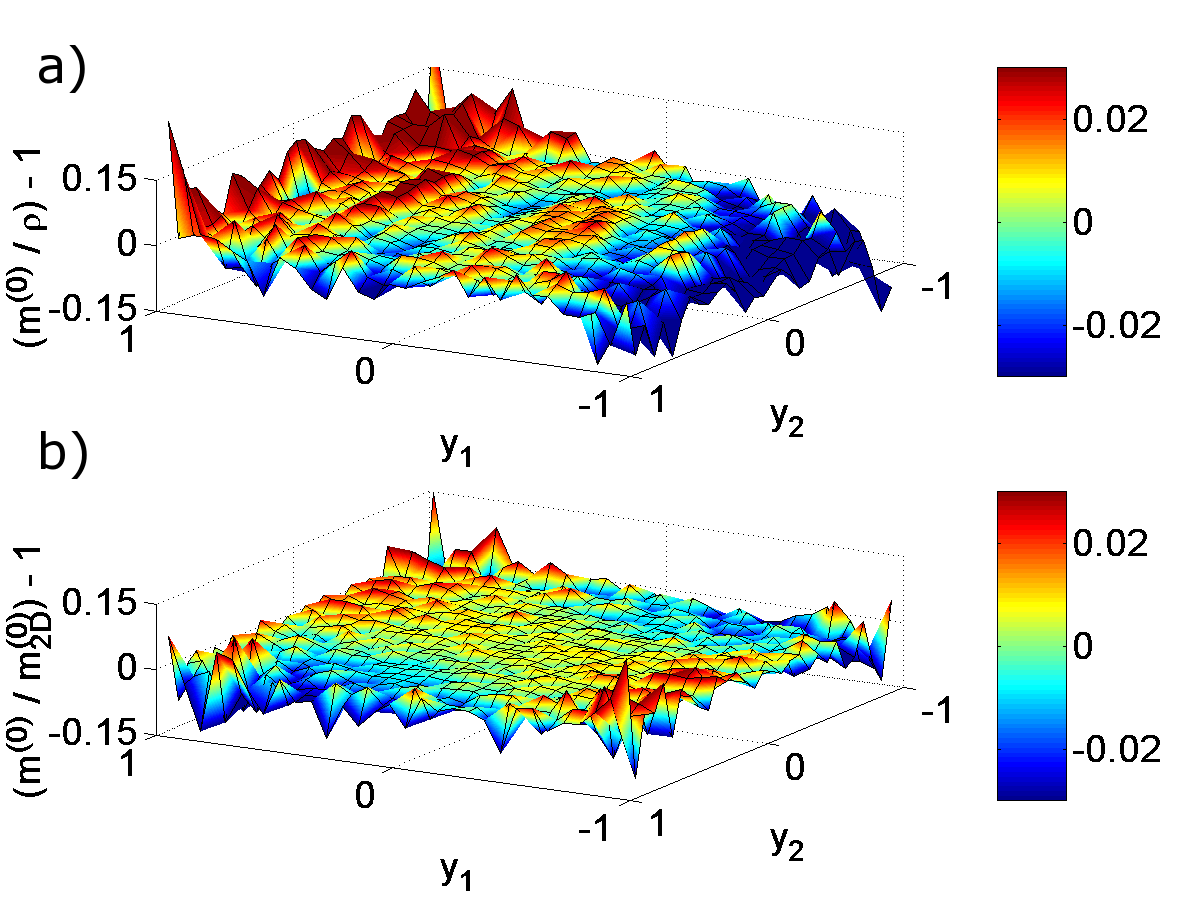}\vspace{-2em}
\end{center}
\caption{ Relative errors of the estimated density values, obtained by a SEA. In ({\bf a}) errors relative
to the true density $\rho$ are shown. In ({\bf b}) errors are relative to the binned density of the $2D$ series.}
\label{fig8}
\end{figure}

Using a SEA also affects the estimates for the process density $\rho$. It would be missleading,
however to compare the estimates $\mom{0}$ to the true density $\rho$, as is done in Fig.~(\ref{fig8}a). To a large extent
the observed errors are caused by finite size effects and not by the reconstruction approach (the binned density of the
true $2D$ series would show very similar errors). To assess the errors that are introduced by the embedding approach,
the estimates $\mom{0}$ thus should be compared to the binned density of the $2D$ series. This is done in
Fig.~(\ref{fig8}b), where the erros are found to be biased by a hyperbolic function in $y_1$ and $y_2$.

Now a MEA, as proposed in Sec.~(\ref{sec_alternative_approach}), is applied. Again the analysis
is purely based on the $1D$ series of positions $Y_1$. Opposed to a SEA, however, velocities are no longer estimated by
a differencing scheme with a fixed step size. Instead, velocities and velocity increments for time increment $\tau$ are
estimated using the step size $\theta\!=\!\tau$. Using a binning approach,
it is not much more effort than for a SEA to implement the calculation of the
density $\mom{0}$ and of the conditional moments $\mom{1}_1/\mom{0}$ and $\mom{2}_{22}/\mom{0}$.
In pseudo-code this reads:
{
\footnotesize
\begin{verbatim}
for i=1:n-kmax  % n data-points
  for k=1:kmax  % kmax increments
    pos = x[i]  % x is data array
    velo = (x[i+k]-x[i])/k/dt  % sampling step dt
    dvelo = (x[i+2*k]-2*x[i+k]+x[i])/k/dt
    idx = getBinIndex(pos,velo)
    if(isValid(idx))
      m0[idx][k] += 1
      m1[idx][k] += dvelo
      m2[idx][k] += dvelo*dvelo
    end
  end
end
for idx=1:idxmax  % loop over all indicees
  for k=1:kmax
    m1[idx][k] /= m0[idx][k] % 1st cond. moment
    m2[idx][k] /= m0[idx][k] % 2nd cond. moment
    m0[idx][k] /= (n-kmax)*binSize % density
  end
end
\end{verbatim}
}

\begin{figure}[h]
\begin{center}
\includegraphics*[width=7.5cm,angle=0]{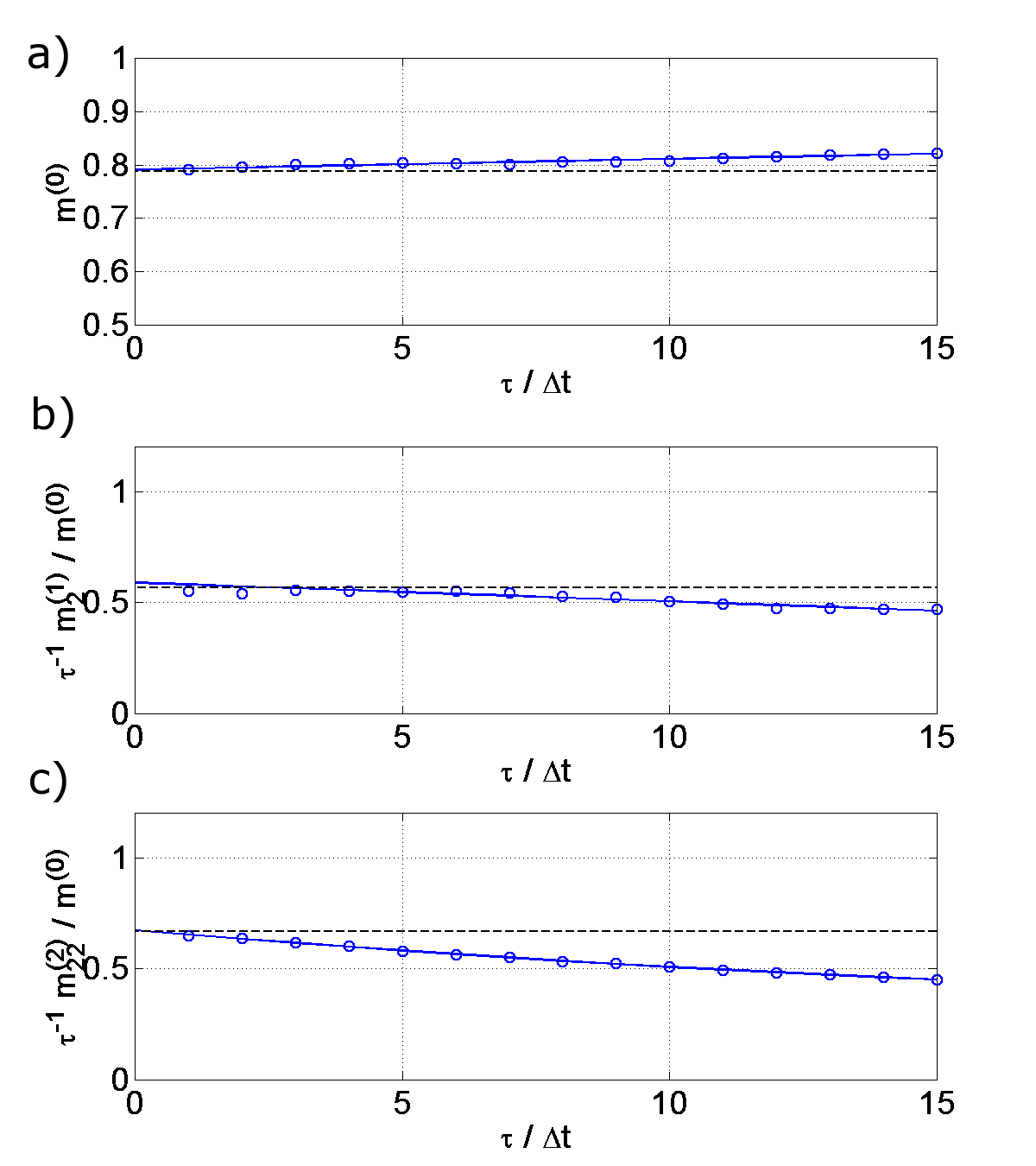}\vspace{-2em}
\end{center}
\caption{ Estimated densities ({\bf a}) and estimated moments (scaled by $\tau^{-1}$) of the conditional velocity increments
({\bf b}, {\bf c}). The estimates (circles) have been obtained by a MEA. The corresponding
fits are shown as solid curves. Estimates are taken at $(y_1,y_2)\!=\!(-0.1,-0.2333)$. Here $\tilde f$
(see Eq.~(\ref{correction_f})) and $2g^2\!/3$ have values of $0.5667$ and $0.6667$ respectively and the binned density of the
$2D$ series has a value of $0.7873$ (dashed lines).}
\label{fig9}
\end{figure}

\noindent Estimates for density and conditional moments obtained by a MEA are shown in Fig.~(\ref{fig9}). As expected from
Eq.~(\ref{relation3_hatm_D}), the scaled moments now approach $\tilde f$ and $2g^2\!/3$ respectively for $\tau\!\to\!0$.
Also the density $\mom{0}$ now depends on $\tau$ and approaches the density of the $2D$ series.

\begin{figure}[h]
\begin{center}
\includegraphics*[width=8.6cm,angle=0]{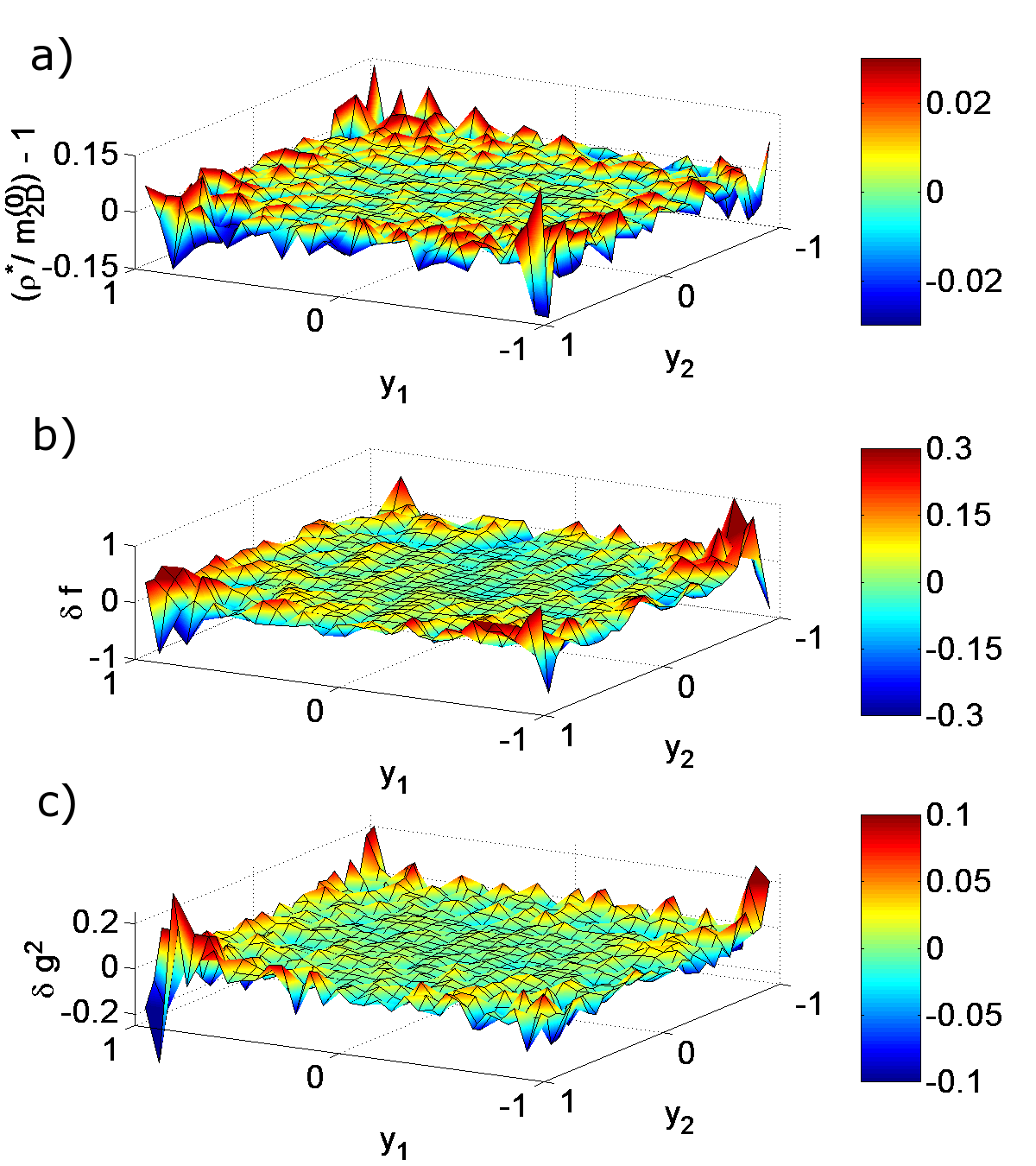}\vspace{-2em}
\end{center}
\caption{ Relative error of the estimated density $\rho$ ({\bf a}) and absolute errors of the estimates for $f$
and $g^2$ ({\bf b}, {\bf c}), obtained by a MEA. Errors in ({\bf a}) are
relative to the binned density of the $2D$ series.}
\label{fig10}
\end{figure}

All increments up to $\tau_\text{max}\!=\!15\Delta t$ are used for the least square fits. The regression functions
$\{1,\tau\}$ are used to fit the density estimates. For the fits of the estimated first and second conditional
moments again the functions $\{\tau,\tau^2\}$ respectively $\{\tau,\tau^2,\tau^3\}$ are used. There is no need to
add functions like $\tau^{-2}$, as still a case without measurement noise is looked at. The errors of the resulting
estimates for $\rho$, $f$ and $g^2$ are shown in Fig.~(\ref{fig10}). Opposed to a SEA, shown in
Fig.~(\ref{fig7}), no obvious biasing of the estimates can be observed and the fluctuations of $\delta f$ and $\delta g^2$
are compareable to those observed in an analysis of the $2D$ series using a SMA.

\section{Embedding approaches with measurement noise}
\label{sec_comparison_noise}

\noindent So far, only data without measurement noise as been analyzed. Next, a series of 'noisy' values $Y_1^*$ is generated
by adding Gaussian, uncorrelated noise with a variance of $V\!=\!1.6667\times10^{-7}$ (this corresponds to a
noise-to-signal amplitude ratio of $10^{-3}$) to the series $Y_1$. This noisy series $Y_1^*$ is then analyzed --
first by applying a SEA and next by applying a MEA. 

\begin{figure}[h]
\begin{center}
\includegraphics*[width=7.5cm,angle=0]{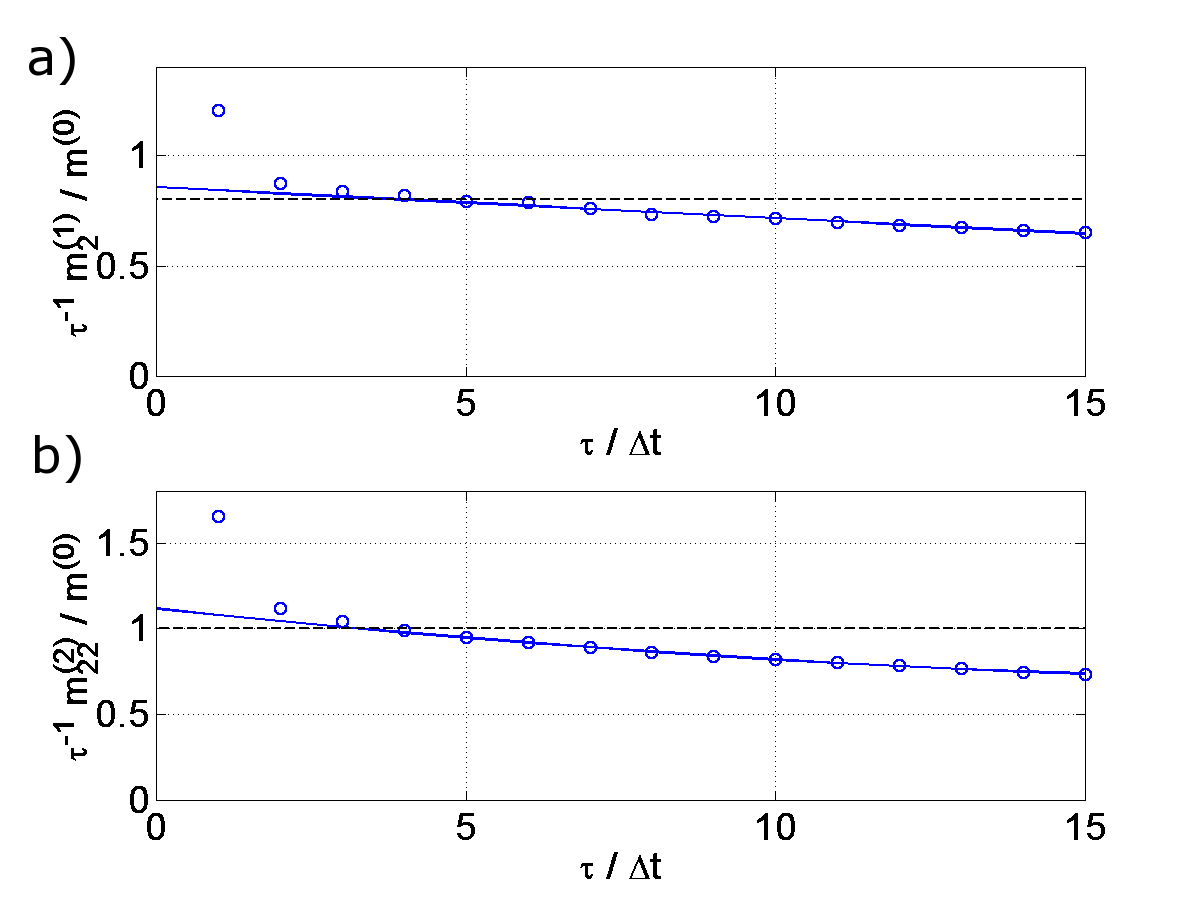}\vspace{-2em}
\end{center}
\caption{ First ({\bf a}) and second moment ({\bf b}) of the conditional velocity increments of the noisy
series $Y_1^*$ (obtained by a SEA with $\theta\!=\!\Delta t$). The estimated values (circles) are scaled
by $\tau^{-1}$. The corresponding polynomial fits are shown as solid curves.
Estimates are taken at $(y_1,y_2)\!=\!(-0.1,-0.2333)$. Here $f$ and
$g^2$ have values of $0.8$ and $1.0$ respectively (dashed lines).}
\label{fig11}
\end{figure}

\begin{figure}[h]
\begin{center}
\includegraphics*[width=8.6cm,angle=0]{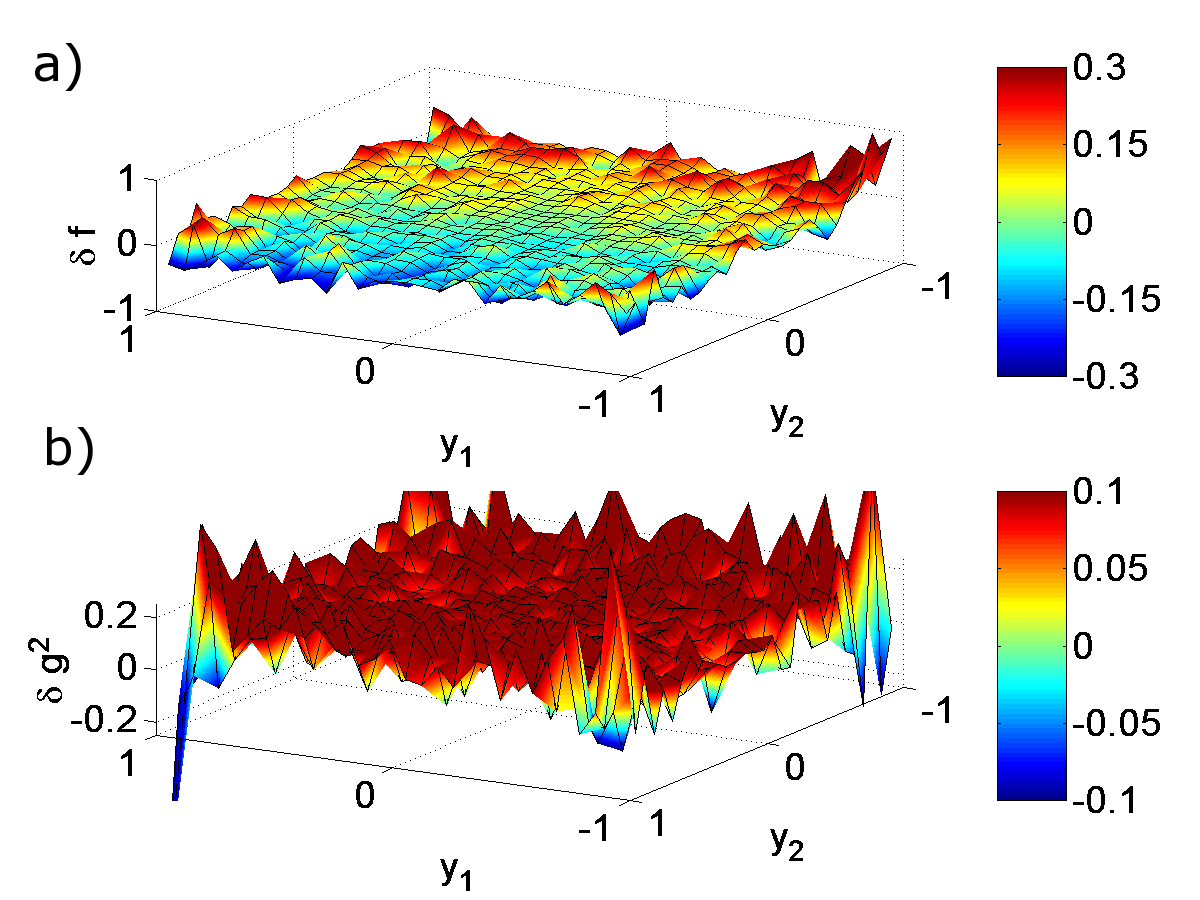}\vspace{-2em}
\end{center}
\caption{ Absolute errors of the estimates for $f$ and $g^2$ ({\bf a},{\bf b}), obtained by
using a SEA for the noisy series $Y_1^*$.}
\label{fig12}
\end{figure}

Moments obtained by a SEA are shown in Fig.~(\ref{fig11}). Due to the measurement noise the scaled moments now
diverge for $\tau\!\to\! 0$. For an estimation of $f$ and $g^2$, therefore, again only increments with
$5 \le\tau/\Delta t \le 15$ are used. Least square fits again are performed using the functions
$\{\tau,\tau^2\}$ and $\{\tau,\tau^2,\tau^3\}$ respectively. The absolute errors $\delta f$ and $\delta g^2$ of the
resulting estimates are shown in Fig.~(\ref{fig12}).
It turns out that $g^2$ is systematically {\em over}-estimated now. The estimates for $f$ still show a significant
bias that is approximately linear in $y_2$ -- allthough the bias now has switched sign.

\begin{figure}[h]
\begin{center}
\includegraphics*[width=7.5cm,angle=0]{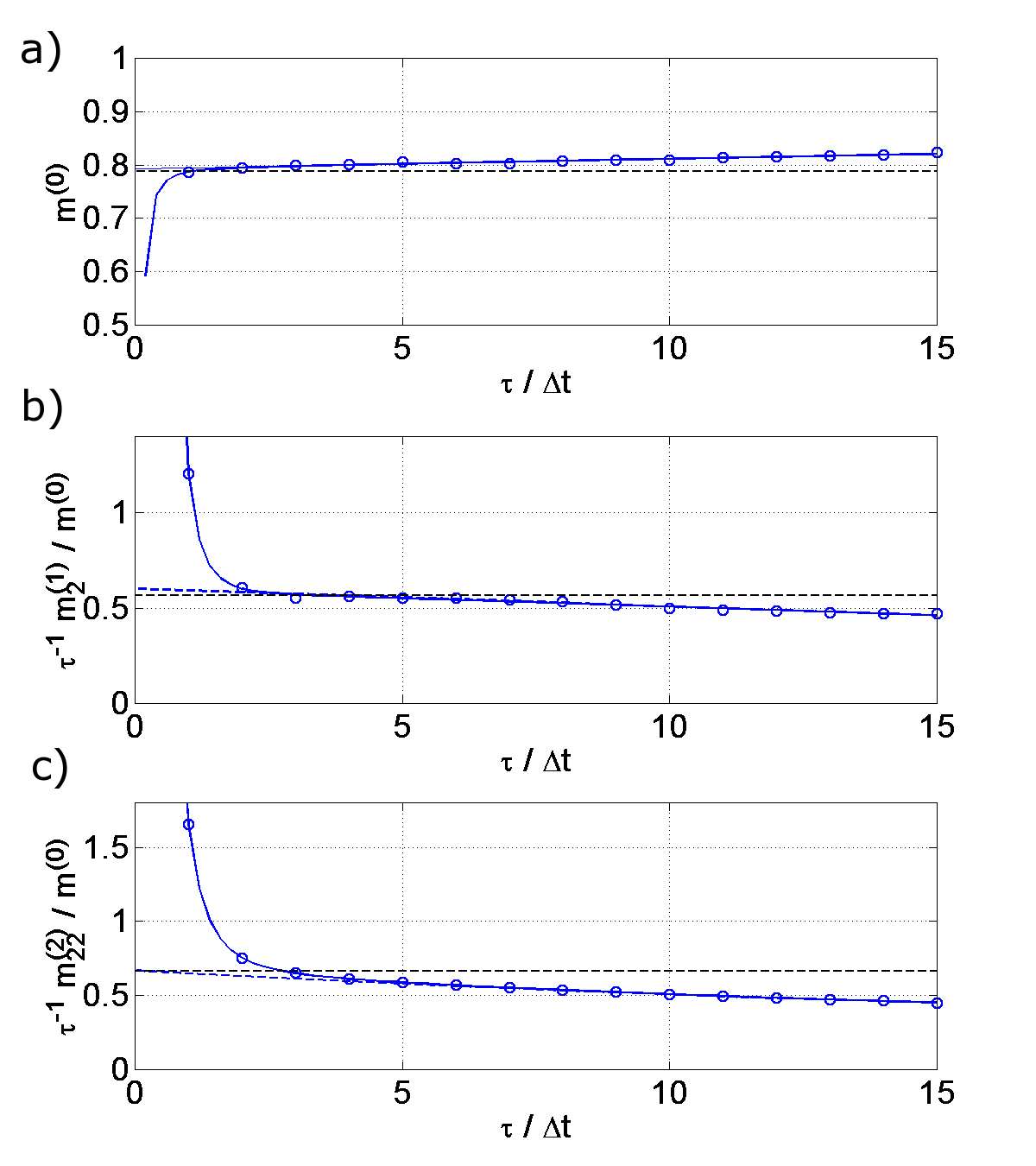}\vspace{-2em}
\end{center}
\caption{ Estimated densities ({\bf a}) and estimated moments (scaled by $\tau^{-1}$) of the conditional velocity increments
({\bf b}, {\bf c}) of the noisy series $Y_1^*$. The estimates (circles) have been obtained by a MEA.
The corresponding fits are shown as solid curves. The non-diverging parts of these fits are shown as dashed curves.
Estimates are taken at $(y_1,y_2)\!=\!(-0.1,-0.2333)$. Here $\tilde f$
(see Eq.~(\ref{correction_f})) and $2g^2\!/3$ have values of $0.5667$ and $0.6667$ respectively and the binned density of the
$2D$ series has a value of $0.7873$ (dashed lines).}
\label{fig13}
\end{figure}

Finally our proposed MEA is applied to the series $Y_1^*$, what leads to estimates for density
and conditional moments as shown in Fig.~(\ref{fig13}). For $\tau\!\to\! 0$ the moments are diverging because
of the terms proportional to $\tau^{-2}$ (and other higher order terms proportional to negative powers of $\tau$), as
described by Eq.~(\ref{relation3_hatm_D}). It is thus neccessary now, to add appropriate regression functions that
account for these terms: For density estimation, all terms up to order $O(\eps)$ are accounted for by using the functions
$\{1,\tau,\tau^{-2}\}$. Fits of the first conditional moments (yielding an estimate for $\tilde f$) are
performed using the regression functions $\{\tau,\tau^{-2},\tau^2,\tau^{-1},\tau^{-4}\}$, i.e. considering terms
up to order $O(\eps^2)$.  Fits of the second conditional moments, finally, (yielding an estimate for $2g^2\!/3$) are
performed using the regression functions $\{\tau,\tau^2,\tau^{-1},\tau^{-4},\tau^3\}$. This choice needs some
explanation. Firstly, only $\tau^3$ is present to account for third order terms. This is a compromise for numerical reasons --
it reduces the number of regression functions and avoids numerical problems with large negative powers of $\tau$.
Secondly, the first order term $6V/\tau^2$ is not accounted for by any regression function. This term is assumed to
be known and thus does not need to be estimated. The value of $V$ is estimated in advance by extrapolating the
auto-covariance function of $Y_1^*$ to $\tau\!=\!0$ and then taking the difference to $\left<Y_1^{*2}\right>$. Estimates for
$V$ that are obtained this way are accurate within about five percent, as has been checked numerically.

\begin{figure}[h]
\begin{center}
\includegraphics*[width=8.6cm,angle=0]{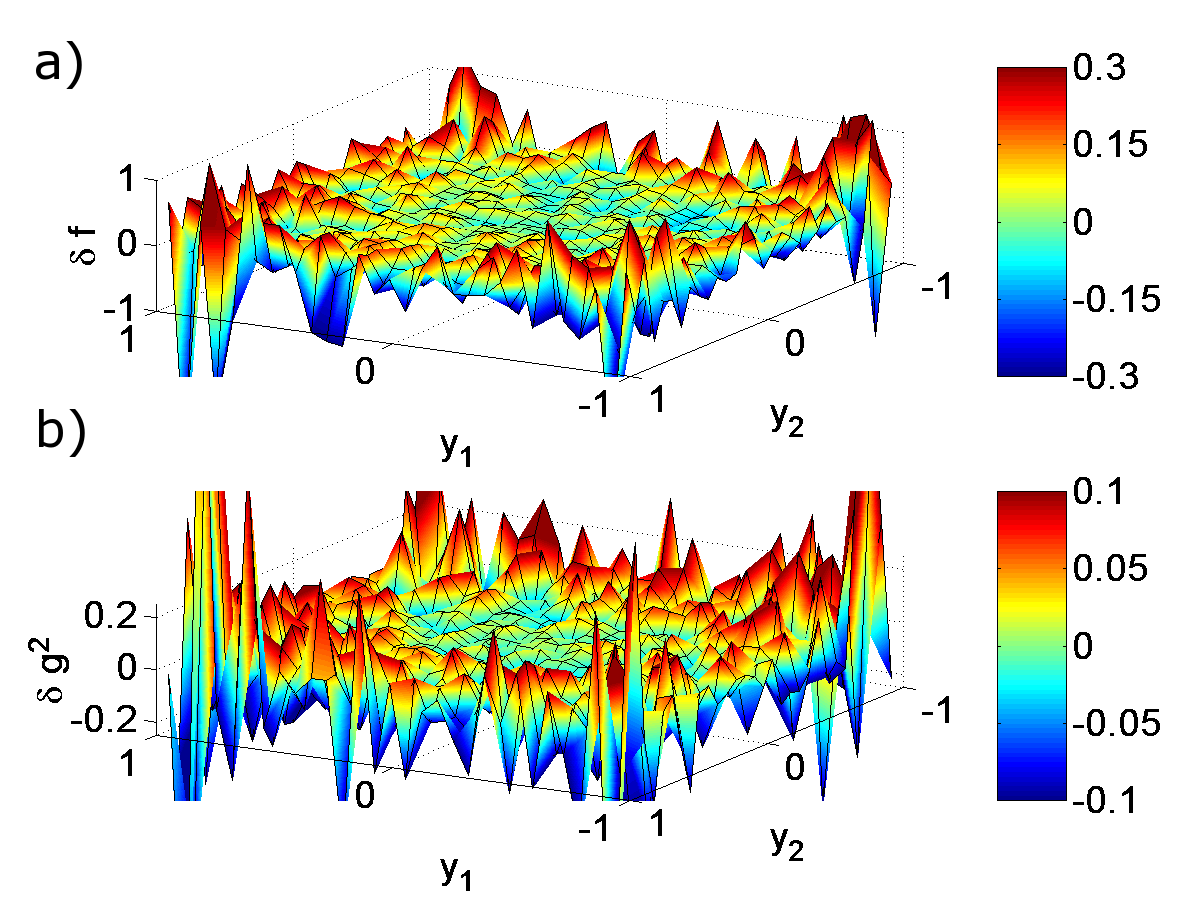}\vspace{-2em}
\end{center}
\caption{ Absolute errors of the estimates for $f$ and $g^2$ ({\bf a}, {\bf b}), obtained by
using a MEA for the noisy series $Y_1^*$.}
\label{fig14}
\end{figure}

Using above sets of regression functions and all increments up to $\tau_\text{max}\!=\!15\Delta t$ then leads to estimates
for $f$ and $g^2$, the absolute errors of which are shown in Fig.~(\ref{fig14}). The estimates are quite heavily
fluctuating now. But -- at least to the bare eye -- the results seem not to be biased.

\section{Summary of numerical results}
\label{sec_summary}

\noindent Solid quantitative results for biases of the results of the different analyses that have been performed would
require an averaging over a large number of analyses of independent realizations of $\vecY$. Instead, a simpler approach
is chosen to numerically compare the results. A polynomial $P$ with
\begin{eqnarray}
P = a\!+\!b_1y_1\!+\!b_2y_2\!+\!c_{11}y_1^2\!+\!c_{12}y_1y_2\!+\!c_{22}y_2^2
\end{eqnarray}

\noindent is fitted to the results for $f$ and $g^2$ using a density weighted least square fit. According to
Eq.~(\ref{def_exp_f_g}) the only non-zero coefficients for $f$ should be $b_1\!=\!-1$ and $b_2\!=\!-3$. For $g^2$
only $a\!=\!1$ should be non-zero. Defining $rms$ as the root of the density weighted mean of the squared differences
between the actual estimates and $P$, allows to also assess the fluctuations. The results for $f$ and $g^2$ are given
in Table~\ref{tab1}.

\begin{table}[h]
\begin{tabular}{| l | *{6}{D{.}{.}{2}} | D{.}{.}{2.2} |}
\cline{2-8}
\multicolumn{1}{c|}{$f$}
               &      a&    b_1& b_2               & c_{11}& c_{12}& c_{22}  & \multicolumn{1}{c|}{$rms\phantom{\!\!\!I^{\int}}$} \\
\hline
exact          &  0.00 & -1.00 & -3.00             &  0.00 &  0.00 &  0.00   & \;\phantom{I^I}  \\
\hline
SMA            &  0.00 & -1.00 & -2.97             & -0.05 &  0.01 &  0.02   & \;0.06\phantom{I^I}  \\
\hline
SEA            &  0.00 & -1.00 & \mbf{-2}.\mbf{77} & -0.05 &  0.01 &  0.02   & \;0.06\phantom{I^I}  \\
MEA            &  0.00 & -0.99 & -2.97             & -0.05 &  0.01 &  0.02   & \;0.06\phantom{I^I}  \\
\hline
$\text{SEA}^*$ &  0.00 & -1.00 & \mbf{-3}.\mbf{16} & -0.05 &  0.01 &  0.01   & \;0.06\phantom{I^I}  \\
$\text{MEA}^*$ &  0.00 & -0.97 & -2.98             & -0.05 &  0.00 &  0.00   & \;\mbf{0}.\mbf{13}\phantom{I^I}  \\
\hline\noalign{\smallskip\smallskip}
\cline{2-8}
\multicolumn{1}{c|}{$g^2$}
               &    a              &    b_1&    b_2& c_{11}& c_{12}& c_{22}  & \multicolumn{1}{c|}{$rms\phantom{\!\!\!I^{\int}}$} \\
\hline
exact          &  1.00             &  0.00 &  0.00 &  0.00 &  0.00 &  0.00   & \;\phantom{I^I}  \\
\hline
SMA            &  1.00             &  0.00 &  0.00 &  0.00 &  0.00 &  0.01   & \;0.02\phantom{I^I}  \\
\hline
SEA            &  \mbf{0}.\mbf{89} &  0.00 &  0.00 &  0.00 &  0.00 & -0.04   & \;0.04\phantom{I^I}  \\
MEA            &  1.00             &  0.00 &  0.00 &  0.00 &  0.01 &  0.00   & \;0.02\phantom{I^I}  \\
\hline
$\text{SEA}^*$ &  \mbf{1}.\mbf{11} &  0.00 &  0.00 &  0.00 &  0.04 &  0.08   & \;0.04\phantom{I^I}  \\
$\text{MEA}^*$ &  1.00             &  0.01 &  0.00 &  0.01 &  0.02 &  0.00   & \;\mbf{0}.\mbf{06}\phantom{I^I}  \\
\hline
\end{tabular}
\caption{Polynomial coefficients and mean errors of a fit of the estimates for $f$ and $g^2$ respectively.
Here $\text{SEA}^*$ and $\text{MEA}^*$ denote results for the noisy series $Y_1^*$. Bold values are discussed in the text.}
\label{tab1}
\end{table}

\noindent The most pronounced effects of a SEA can be observed for the coefficient $b_2$, when estimating
$f$, respectively for the coefficient $a$, when estimating $g^2$. These coefficients are also strongest affected by the
presence of measurement noise. Applying a MEA, however, yields results that are compareable to
those obtained by an analysis of the $2D$ series -- at least if no measurement noise is present. For noisy data the
coefficients still are quite accurate but the mean error, $rms$, becomes larger then. This is a consequence of the
large number of regression functions that is required for the analysis of noisy data. 

\section{Conclusions}
\label{app_conclusions}

\noindent For a time series analysis of a process $\vecX$ that is described by a stochastically forced second order ODE,
frequently an embedding strategy as outlined in Sec.~\ref{sec_intro} is used: First the temporal derivative $\dot\vecX$ is
estimated for each point in time by a numerical differencing scheme, and a new series $\vecY^t\!:=\!(\vecX^t,\dot\vecX^t)$
is built. Then a Markov analysis is applied to the series $\vecY$ in order to estimate its drift- and diffusion
functions. However, the errors that are caused by the differencing scheme lead to notably biased estimates for
these functions. Additionally, even a very small amount of measurement noise
has strong influence on the results.

The errors of the above 'standard' approach have been studied analytically and a modified approach has been proposed.
This approach allows for an accurate estimation of the drift- and diffusion functions and, additionally, is able to
deal with weak measurement noise. This has been verified for a numerical test case.

In this numerical test it also could be seen that measurement noise is a bigger problem than one might think
intuitively.
Already measurement noise with an noise-to-signal amplitude ratio of $10^{-3}$ had a severe influence:
For the standard approach, it introduces an additional, notable bias to the results. For the modified approach,
however, the results stay unbiased. Here the presence of noise only affects the fluctuations, which
become much stronger.
 
The implementation of the presented approach is easily done and straight forward. The algorithm is not demanding with
respect to memory or CPU power. All calculations have been performed on a standard desktop PC, where each analysis took
less than one minute.

Compared to the standard approach, our modified embedding approach performs much better at compareable costs.
It, therefore, should be the method of choice in the given setup.

\appendix
\section{Taylor--\Ito\ expansion of $\vecY_1$}
\label{app_taylor_ito}

\noindent A Taylor--\Ito\ expansion of $\vecY_1(t)$ provides a stochastic description of the values $\vecY_1(t\!+\!\Delta)$
for given $\vecY(t)$. Assuming smooth functions $\vecf$ and $\matg$, the expansion can be written as an infinite sum of
deterministic and stochastic integrals that only depend on $\Delta$ and $\vecxi$ and that are weighted by
coefficient functions. These functions only depend on the values and derivatives of $\vecf$ and $\matg$, evaluated at $\vecY(t)$.
In the following, some properties of the integrals will shortly be summarized. A detailed description of the Taylor--\Ito\ 
expansion and the properties of the stochastic integrals can be found e.g. in \cite{platen99}.

Using a multi-index $\vecalpha$, the expansion of $\vecY_1$ can be written quite compactly
\begin{eqnarray}
\vecY_1(t\!+\!\Delta)\big|_\vecy &=& \vecy_1+\sum_\vecalpha \vecc_\vecalpha(\vecy)I_\vecalpha^{t,\Delta}(\vecxi),
\end{eqnarray}

\noindent with
\begin{eqnarray}
\vecalpha &:=& (\alpha_1,\ldots,\alpha_n),\quad n\in{\mathbb N},\\
\alpha_i &\in& \{0,\ldots,N\}.
\end{eqnarray}

\noindent Here $\vecc_\vecalpha$ denotes the above mentioned coefficient functions. The multiple integrals $I_\vecalpha$
may contain integrations with respect to time as well as integrations with respect to components of the Wiener process
$\mbf{W}(t)$, associated with the Gaussian noise $\vecxi(t)$. The structure of each integral is determined by its
multi-index $\vecalpha$
\begin{eqnarray}
I^{t,\Delta}_\vecalpha &:=&
\int_{s_n=t}^{t+\Delta}
\int_{s_{n\!-\!1}=t}^{s_n}\!\cdots
\int_{s_1=t}^{s_2}
\,dZ_1\cdots dZ_n
,
\end{eqnarray}

\noindent with
\begin{eqnarray}
dZ_i &:=& \left\{\begin{array}{ll}
ds_i & \quad, \alpha_i=0 \\
dW_{\!\alpha_i}(s_i) & \quad, \alpha_i\ne 0
\end{array}\right. .
\end{eqnarray}

\noindent The multi-index also determines the order of magnitude of the integral
\begin{eqnarray}
I^{t,\Delta}_\vecalpha &=& O(\Delta^{m(\vecalpha)}),
\end{eqnarray}

\noindent with
\begin{eqnarray}\label{def_m_alpha}
m(\vecalpha) &:=& \sum_{\alpha_i=0} 1 + \sum_{\alpha_i\ne0} \frac{1}{2}.
\end{eqnarray}

\noindent Because of \Ito's definition of the stochastic integral, the expectation value of $I_\vecalpha$ will be
zero if it contains any integration with respect to a Wiener process, i.e. if there are any non-zero components in
it's index-vector. Otherwise, when all components are zero, the integral becomes purely deterministic and evaluates to
$(\Delta^{\!n})/n!$, where $n$ indicates the length of $\vecalpha$.

In App.~\ref{app_higher_order} expectation values of multiple products of integrals will be of interest. This will be
restricted to such products, however, where the increments $\Delta_i$ of the integrals are all of the same order of magnitude
\begin{eqnarray}
\left<\prod_{i=1}^k I^{t,\Delta_i}_{\vecalpha_i}\right> &=&
\left\{\begin{array}{l}
O(\Delta^{\!r}) \\
0
\end{array}\right.,
\end{eqnarray}

\noindent with
\begin{eqnarray}
\Delta_i &\overset{!}{=}& O(\Delta),\quad
r \;:=\;\sum_{i=1}^k \!m(\vecalpha_i).
\end{eqnarray}

\noindent Here a sufficient (but not neccessary) condition for a vanishing expectation value is an odd total number of
non-zero entries in the index vectors, i.e. a non-integral value of $r$. Non-vanishing expectation
values will thus always have the magnitude of an integral power of $\Delta$.
For an even more restrictive case, where the ratios $\Delta_i/\Delta_j$ of the increments are kept fix, the expectation value
actually becomes proportional to a power of $\Delta$ (this can be shown by scaling the time variables in the integrals and
using the fact that $\lambda^{1/2}\vecxi(\lambda t)$ is (statistically) identical to $\vecxi(t)$)
\begin{eqnarray}
\left<\prod_{i=1}^k I^{t,\Delta_i}_{\vecalpha_i}\right> &=&
\left\{\begin{array}{l}
C \cdot\Delta^{\!r} \\
0
\end{array}\right.,
\end{eqnarray}

\noindent with
\begin{eqnarray}
\Delta_i &\overset{!}{=}& \lambda_i\Delta\, , \quad
\lambda_i \;=\; \text{const} \;\overset{!}{=} O(1).
\end{eqnarray}

\noindent This only holds for $\lambda_i\!=\!\text{const}$. Otherwise the expectation values will in general not
have a uniform definition but will be given by multivariate polynomials in the variables $\lambda_i\Delta$ with
coefficients depending on size relations of the increments. The following explicit expectation value may serve as
an example, but the result is also actually used in the calculation of the moments $\matM^{(k,\nu)}$ in
Sec.~\ref{sec_moments_Mknu}
\begin{eqnarray}\label{def_var_Ii0}
&&\!\!\!\!\!\!\!\left<I_{(i,0)}^{t,\Delta_1}I_{(j,0)}^{t,\Delta_2}\right> = \cr
&&\qquad\quad\delta_{ij}\left\{\begin{array}{ll}
\frac{1}{2}\Delta_1\Delta_2^2-\frac{1}{6}\Delta_2^3\;, & \Delta_2 \le \Delta_1 \\[.5em]
\frac{1}{2}\Delta_1^2\Delta_2-\frac{1}{6}\Delta_1^3\;, & \Delta_2 > \Delta_1
\end{array}\right. .
\end{eqnarray}

\noindent Next the actual expansion will be given. There is one special point in the expansion of $\vecY_1$: If the
last entry of an index-vector is non-zero, the corresponding coefficient function $\vecc_\vecalpha$ will be vanishing
(this is due to the fact that $\vecY_1$ is not directly driven by noise; see Eq.~(\ref{Langevin_Yemb})).
The remaining integrals will thus all be at least of order $O(\Delta)$
\begin{eqnarray}\label{expansion_Y1}
\vecY_1(t\!+\!\Delta)\big|_\vecy
 &=& \vecy_1+\vecy_2\,\Delta+\vecf(\vecy)\frac{\Delta^2}{2}
    +\matg(\vecy)\mbf{I}^{t,\Delta}\cr
&&+\vecR^{t,\Delta}\!(\vecy),
\end{eqnarray}

\noindent with
\begin{eqnarray}
I_i^{t,\Delta} &:=& I_{(i,0)}^{t,\Delta}.
\end{eqnarray}

\noindent The remainder $\mbf{R}$ is used to summarize all remaining expansion terms. Its
lowest order stochastic terms are given by $\vecc_{(j,k,0)}I_{(j,k,0)}^{t,\Delta}$ and its lowest order deterministic
term by $\vecc_{(0,0,0)}I_{(0,0,0)}^{t,\Delta}$. Thus $\mbf{R}$ is a term of order $O(\Delta^2)$ with the statistical
properties
\begin{eqnarray}
\left<R_i^{t,\Delta}\right> &=& O(\Delta^3),\\
\left<R_i^{t,\Delta}R_j^{t,\Delta}\right> &=& O(\Delta^4).
\end{eqnarray}

\section{Functional form of higher order terms}
\label{app_higher_order}

\noindent Equation (\ref{relation_mstar_D}) is accurate up to first order only. The 'classical' Markov analysis, as
sketched in Sec.~\ref{sec_intro}, faces the same problem: Equation~(\ref{relation_m_D}) the relation between the moments
$\bmom{k}$ and the Kramers--Moyal coefficients, is accurate up to order $O(\tau)$ only. However, for
Eq.~(\ref{relation_m_D}) the functional form (with respect to $\tau$) of the higher order terms is known -- terms of order
$O(\tau^n)$ simply are proportional to $\tau^n$. Performing a linear regression with a function-base
$\{\tau,\tau^2,\ldots,\tau^n\}$ will thus allow parameter estimations with an accuracy of $O(\tau^n)$ (of cause, there
are practical limitations for $n$).

For higher order estimations in the given setup, the functional form (with respect to $\tau$ and $\theta$) of the
higher order terms of $\bhmom{k}$ is needed. Because the functional form of {\em all} terms of $\bhmom{k}$ is dictated by the
form of the moments $\matM^{(k,\nu)}\!=\!\left<\mbf{A}^k\otimes\mbf{B}^\nu\right>$, the starting point will be the
vectors $\mbf{A}$ and $\mbf{B}$.

According to Eq.~(\ref{def_A_B_explicit}) the components of both vectors can be expressed as linear combination of
terms that either stem from the Taylor--\Ito\ expansion or from the measurement noise. Denoting the former by
$q^\xi$ and the later by $q^\gamma$, the terms can be expressed as (using $m(\vecalpha)$ as defined in
Eq.~(\ref{def_m_alpha}))
\begin{eqnarray}
q^\xi &\in& \{ I_\vecalpha^{t,\Delta}, \theta^{-1}I_\vecbeta^{t,\Delta} \},\\
\Delta &\in& \{\theta,\tau,\tau\!+\!\theta \},\;\;
m(\vecalpha) \ge 1,\;\;
m(\vecbeta) \ge 3/2,
\end{eqnarray}

\noindent and
\begin{eqnarray}
q^\gamma &\in& \{ \Gamma_i(t\!+\!\Delta), \theta^{-1}\Gamma_i(t\!+\!\Delta) \},\\
\Delta &\in& \{0,\theta,\tau,\tau\!+\!\theta \}.
\end{eqnarray}

\noindent A component of $\matM^{(k,\nu)}$, therefore, can be expressed as a linear combination of expectation values
of $k\!+\!\nu$ factors $q$. Because $\vecGamma$ is assumed to be external noise, each
expectation value, denoted by $Q$, can be factorized.
\begin{eqnarray}
Q &:=& \left< \prod_{i=1}^{n_1}q_i^\xi \prod_{j=1}^{n_2}q_j^\gamma \right> =
 \left< \prod_{i=1}^{n_1}q_i^\xi \right> \left< \prod_{j=1}^{n_2}q_j^\gamma \right>,
\end{eqnarray}

\noindent with
\begin{eqnarray}
n_1\!+\!n_2 &=& k\!+\!\nu.
\end{eqnarray}

\noindent The components $\Gamma_i$ have been assumed to be Gaussian noise with a magnitude of $O(\eps^{3/2})$.
A non-vanishing expectation value of a product of $n$ factors $\Gamma_i(t\!+\!\Delta_i)$ will thus be given by
$C^\gamma\eps^{3n/2}$, where $C^\gamma$ in general depends on whether $\tau$
equals $\theta$ or not. As $q^\gamma$ either denotes a factor $\Gamma$ or a factor $\theta^{-1}\Gamma$, one finds
\begin{eqnarray}
\left< \prod_{j=1}^{n_2}q_j^\gamma \right> &=&
C^\gamma \theta^{n_2\!-\!n_2'} \left(\eps^3/\theta^2\right)^{n_2/2},
\end{eqnarray}

\noindent with
\begin{eqnarray}
0 \le n_2' \le n_2.
\end{eqnarray}

\noindent The expectation value of a product of integrals $I_\vecalpha$ will be a polynomial $P$ in $\tau$ and $\theta$,
where the coefficients in general will depend on whether $\tau$ is smaller than $\theta$ or not. For each monomial the powers
of $\tau$ and $\theta$ will sum up to a value $n_1''$, determined by the index-vectors of the integrals.
 As $q^\xi$ either denotes a factor $I_\vecalpha$ or a factor $\theta^{-1}I_\vecbeta$, one finds
\begin{eqnarray}
\left< \prod_{i=1}^{n_1}q_i^\xi \right> &=&
\theta^{-n_1'}P^{(n_1'')}(\tau,\theta),
\end{eqnarray}

\noindent with
\begin{eqnarray}
0 \le n_1' \le n_1,\quad n_1'' \ge n_1+n_1'/2.
\end{eqnarray}

\noindent The expectation values $Q$ can thus be written as a linear
combination of terms $Q'$, as defined below. Here it has been used that odd moments of $\Gamma$ are vanishing, i.e. only
even values $n_2$ have to be considered.
\begin{eqnarray}
Q' &=& C \tau^a \theta^b\left(\eps^3/\theta^2\right)^c,
\end{eqnarray}

\noindent with
\begin{eqnarray}
a \ge 0,\quad a+b \ge 0,\quad c \ge 0.
\end{eqnarray}

\noindent The value of $C$ depends on whether $\tau$ is smaller, equal or larger than $\theta$. Keeping the
ratio of $\tau$ and $\theta$ fix, therefore, leads to a constant factor $C$. The functional form of the  terms $Q'$ (and
thus of all terms in the moments $\bhmom{k}$) is then given by
\begin{eqnarray}
\tau/\theta &\overset{!}{=}& \text{const}\quad\Rightarrow\quad
Q' \;\sim\; \tau^a \left(\eps^3/\tau^2\right)^b,
\end{eqnarray}

\noindent with
\begin{eqnarray}
a \ge 0,\quad b \ge 0.
\end{eqnarray}

\noindent Because $\tau$ is assumed to be of order $O(\eps)$, the term $Q'$ is of order $O(\eps^{a+b})$. The function-base of
terms of order $O(\eps^n)$, denoted by ${\cal B}^{(n)}$, thus consists of the $\tau$-dependend parts of all terms $Q'$ with
$a\!+\!b\!=\!n$

\begin{eqnarray}\label{def_function_bases}
{\cal B}^{(0)} &=& \{ 1 \},\\
{\cal B}^{(1)} &=& \{ \tau, \tau^{-2} \},\\
&\vdots&\\
{\cal B}^{(n)} &=& \{ \tau^n\!, \tau^{n-3}\!,\ldots, \tau^{-2n} \}.
\end{eqnarray}

\noindent Unfortunately this means ${\cal B}^{(n)}\subset{\cal B}^{(n+3)}$, which puts a limit on the accuracy that can
be achieved. It is, for example, not possible to distinct some of the terms of order $O(\eps^4)$ from the terms of order
$O(\eps)$. At most, therefore, an accuracy of order three can be achieved (if no $O(1)$ terms are present).


%
\end{document}